%% file: ActiveQuantumFlocks_arXiv.tex
\documentclass[aps,prx,superscriptaddress,twocolumn,showpacs]{revtex4-1}
\usepackage[table]{xcolor}
\usepackage{pgf}
\usepackage[utf8]{inputenc}
\usepackage{amsmath}
\usepackage{braket}
\usepackage{amssymb}
\usepackage{amsmath}
\usepackage{bbold}
\usepackage{enumitem} 
\usepackage{appendix}
\usepackage{graphicx}
\usepackage{mathtools}
\usepackage{commath}
\usepackage{romannum}
\usepackage{lipsum}
\usepackage{tikz}
\usepackage{bbold}
\usepackage{xfakebold}

\newcommand{\fbseries}{\unskip\setBold\aftergroup\unsetBold\aftergroup\ignorespaces}
\makeatletter
\newcommand{\setBoldness}[1]{\def\fake@bold{#1}}
\makeatother

\usepackage{epstopdf}
\usepackage{amsmath}
\NewDocumentCommand{\tens}{e{_^}}{%
  \mathbin{\mathop{\otimes}\displaylimits
    \IfValueT{#1}{_{#1}}
    \IfValueT{#2}{^{#2}}
  }%
}
\usepackage[pdftex,bookmarks,colorlinks]{hyperref}
\usepackage{hyperref}

\usepackage[normalem]{ulem}
\usepackage{dsfont}

\usepackage{soul}

\newcommand{\up}{\uparrow}
\newcommand{\down}{\downarrow}

\newcommand{\changed}[1]{\textcolor{olive}{#1}}

\usepackage{todonotes}

\usepackage{hyperref}

\begin{document}
\title{Active quantum flocks}
\author{Reyhaneh Khasseh}
\affiliation{Theoretical Physics III, Center for Electronic Correlations and Magnetism,
Institute of Physics, University of Augsburg, D-86135 Augsburg, Germany}
\affiliation{Max-Planck-Institut f\"{u}r Physik komplexer Systeme, 01187 Dresden, Germany}

\author{Sascha Wald}
\affiliation{Statistical Physics Group, Centre for Fluid and Complex Systems,
Coventry University, Coventry, England}

\affiliation{$\mathbb{L}^4$ Collaboration \& Doctoral College for the
Statistical Physics of Complex Systems,
Leipzig-Lorraine-Lviv-Coventry, Europe}

\author{Roderich Moessner}

\affiliation{Max-Planck-Institut f\"{u}r Physik komplexer Systeme, 01187 Dresden, Germany}

\author{Christoph A. Weber}

\affiliation{Faculty of Mathematics, Natural Sciences, and Materials Engineering: Institute of Physics, University of Augsburg, Universit\"atsstra{\ss}e\ 1, 86159 Augsburg, Germany}
\affiliation{Max-Planck-Institut f\"{u}r Physik komplexer Systeme, 01187 Dresden, Germany}

\author{Markus Heyl}

\affiliation{Theoretical Physics III, Center for Electronic Correlations and Magnetism,
Institute of Physics, University of Augsburg, D-86135 Augsburg, Germany}
\affiliation{Max-Planck-Institut f\"{u}r Physik komplexer Systeme, 01187 Dresden, Germany}

\begin{abstract}
%
Flocks of animals represent a prominent archetype of collective behavior in the macroscopic classical world, where the constituents, such as birds, concertedly perform motions and actions as if being one single entity.
Here, we address the so far open question of whether flocks can also form in the microscopic world at the quantum level.
For that purpose, we introduce the concept of \textit{active quantum matter} by formulating a class of models of active quantum particles on a one-dimensional lattice.
We provide both analytical and large-scale numerical evidence that these systems can give rise to quantum flocks.
A key finding is that these quantum flocks exhibit distinct quantum properties by developing strong quantum coherence over long distances.
We propose that quantum flocks could be experimentally observed in Rydberg atom arrays.
%
%
%
\end{abstract}

\maketitle

\textit{Introduction:--} In the quantum world, remarkable advances in quantum simulators have led to unprecedented capabilities in controlling and probing the real-time dynamics of quantum matter~\cite{Gross2017,Monroe2021,Browaeys2020,Blais2021}.
%
Among the most important developments have been the observation of genuinely nonequilibrium phases of matter such as many-body localization~\cite{Schreiber2015,Smith2016}, discrete time crystals~\cite{Choi2017,Zhang2017}, or quantum many-body scars~\cite{Bernien_2017}.
In the classical world, progress in the understanding of dynamical processes maintained away from equilibrium in the context of active matter systems, e.g.\ describing the physics of biological systems, has been distinctly impressive~\cite{julicher1997modeling, ramaswamy2010mechanics,marchetti2013hydrodynamics,prost2015active, cates2018theories, weber2019physics}.
This, in particular, includes one of the most prominent archetypes of collective motion - flocks, which realize an oriented, clustered, and moving collection of constituents~\cite{Vicsek_1995, toner2005hydrodynamics, chate2020dry}.
So far, these developments on the nonequilibrium many-body physics of quantum and classical systems have evolved independently, leaving open the fundamental question emerging naturally at their interface: is it also possible for quantum particles to exhibit flocking,  similar to birds or fish in the classical world?

In this work, we introduce the concept of active quantum matter, by formulating key underlying quantum dynamical processes.
We then provide evidence that these particles can realize a quantum flock in that they form collectively moving clusters which spontaneously undergo polar symmetry breaking. 
We observe that the resulting flock experiences distinct quantum features absent in the classical world.
Specifically, our active quantum system exhibits a pronounced long-distance quantum coherence suggesting coherent, ballistic motion over large distances.
We argue that the identified underlying mechanism is general and can be potentially used to realize also other phases of active quantum matter.
Our work therefore opens up a route towards exploring yet unknown classes of nonequilibrium states in quantum many-body systems with the potential to realize also other collective behaviors of biological active matter systems in a quantum context.
%



\begin{figure*}
	\centering
	\includegraphics[width=\textwidth]{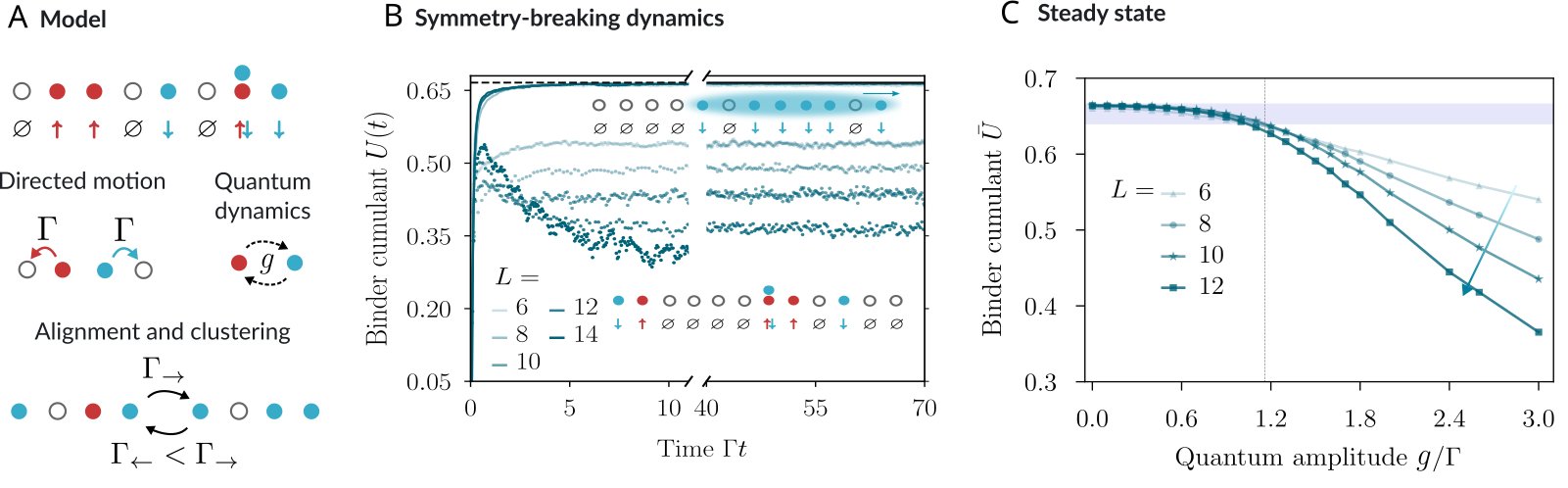}
	\caption{\textbf{Active quantum flocks.} \textbf{A} The model for active quantum matter involves two species of hard-core bosons ($\uparrow,\downarrow)$ on a one-dimensional lattice subject to three types of dynamical processes. The particles perform a dissipative directed motion at a rate $\Gamma$ with $\uparrow$ and $\downarrow$ particles moving to the left and right, respectively. Quantum dynamics is introduced by a coherent spin-flip process occurring with an amplitude $g$. Alignment and clustering of particles is enabled by a conditional dissipative spin-flip process whose rate effectively depends on the surrounding magnetization. Here, the rate $\Gamma_\leftarrow < \Gamma_\rightarrow$ as the environment is dominated by $\downarrow$ particles. \textbf{B}  Dynamics of the Binder cumulant $U(t)$ for different quantum amplitudes $g$ indicating long-range order for weak $g=0.2$ (solid lines), whereas for large $g=3.0$ (dotted lines) the Binder cumulant exhibits a tendency towards a vanishing value upon increasing $L$ suggesting a disordered phase. The numerical data has been obtained for an alignment parameter $K=3.8$ and $N_r=1000$ trajectories. \textbf{C} Long-time average of the Binder cumulant $\overline{U}$ as a function of the quantum amplitude $g$ of the quantum dynamics for different system sizes $L$ at alignment parameter $K=3.8$. For weak $g$ the Binder cumulant approaches a value of $U=2/3$ indicative of a quantum flocking phase, whereas for large $g$ the Binder cumulant displays a tendency towards a vanishing value upon increasing $L$. The horizontal shaded area indicates the threshold used to estimate the phase transition point, which is included as the dashed line.
 }
	\label{fig:aqim}
\end{figure*}

\textit{Model for active quantum flocks:--} It is the key aim of this work to devise a quantum-mechanical analogue of active matter systems~\cite{Solon_2013,Solon_2015}, with the particular goal to realize flocks in the quantum world.
For that purpose, it will be necessary to identify quantum counterparts to the following two processes.
First, it would be key to determine how to make quantum particles active, which in analogy to the classical world would mean that their motion is continuously supplied with energy on the single-particle level from an environment, breaking local detailed balance and enabling persistent motion~\cite{ramaswamy2010mechanics, marchetti2013hydrodynamics,weber2014defect, fodor2022irreversibility, bar2020self}. 
Second, in order for such active particles to form a flock, a so-called alignment process is required describing an interaction which correlates the particles direction of motion according to their environment, such as birds aligning their velocity with their neighbors' in the pioneering Vicsek model~\cite{Vicsek_1995}.
This, however, imposes a direct challenge in the quantum world, since, quantum mechanically, positions and velocities (or momenta) are non-commuting observables.

In the following, we introduce a general class of quantum models which address these challenges.
We consider a system of hard-core bosons on a one-dimensional chain of $L$ lattice sites with periodic boundary conditions and two species of particles labeled by an effective spin $\sigma=\uparrow,\downarrow$, see Fig.~\ref{fig:aqim}A.
In Supplementary Material Sec.~VIII, we discuss how to realize the individual dynamical processes microscopically in systems of Rydberg atoms.

It is key for active matter to include dissipative processes.
We choose environments that can be effectively described by a Lindblad master equation~\cite{gardiner2004quantum}:
\begin{equation}\label{eq:Lindblad}
\frac{d\hat{\rho}}{dt}=-i[\hat{H},\hat{\rho}]+{\cal D}(\hat{\rho})~\, .
\end{equation} 
which appear genuinely when coupling quantum matter to photons.
Here, $\hat \rho$ denotes the density matrix of the quantum system. 
Two types of contributions can be distinguished: the coherent evolution by the Hamiltonian $\hat H$, and the dissipative dynamics generated through $\mathcal{D}(\hat \rho)$.
We now introduce specific environments, which realize the aforementioned two key desired processes responsible, firstly, for the active  motion (labeled by $X=\mathcal{M}$ in the following) and, secondly, for the alignment ($X=\mathcal{A}$).
This leads to a decomposition $\mathcal{D}(\hat \rho) = \mathcal{D}_{\mathcal{M}}(\hat \rho) + \mathcal{D}_{\mathcal{A}}(\hat \rho)$, where ${\cal D}_{ X}(\hat{\rho})=\sum_{l=1}^{L}\sum_{\sigma=\uparrow, \downarrow} (\Gamma_X/2) (2\hat{X}_{l\sigma}\hat{\rho}\hat{X}^{\dagger}_{l\sigma}-\hat{X}^{\dagger}_{l\sigma}\hat{X}_{l\sigma}\hat \rho  - \hat \rho \hat{X}^{\dagger}_{l\sigma}\hat{X}_{l\sigma})$.
Here, $\hat X_{l\sigma}$ denotes the respective quantum jump operator of species $\sigma$ on lattice site $l$ and $\Gamma_X$ is the rate of the corresponding process. An alternative formulation of the dissipative contributions is discussed in Supplementary Material Sec. VI.

In order to make the system active we choose quantum jump operators $\hat{\mathcal{M}}_{l\uparrow}=\hat{c}^{\dagger}_{l\uparrow}\hat{c}_{l+1\uparrow},~\hat{\mathcal{M}}_{l\downarrow}=\hat{c}^{\dagger}_{l+1\downarrow}\hat{c}_{l\downarrow}$.
Here, $\hat c_{l\sigma}^\dag$ denotes the creation operator for a hard-core boson of type $\sigma$ at lattice site $l$.
This contribution leads to a directed motion of spin-$\uparrow$ and $\downarrow$ particles to the left and right, respectively, in close analogy to classical active Ising models~\cite{Solon_2013,Solon_2015}.
Importantly, this dynamical process violates Kolmogorov's criterion implying the breaking of local detailed balance (see Supplementary Material Sec. I).
While recent works already aimed at identifying active quantum processes for single particles~\cite{Zheng2023,Yamagishi2023}, the process presented here induces activity on a many-body level.

In a next step we now aim to address the challenge of realizing a local alignment of velocities.
For that purpose, we introduce a dissipative process aligning locally the internal degree of freedom $\sigma$ of the particles, which in turn also aligns their direction of motion.
Concretely, we choose $\hat{\mathcal{A}}_{l\sigma}=\hat{c}^{\dagger}_{l\sigma}\hat{c}_{l\overline{\sigma}} \hat P_l~$ inducing transitions between the two particle species.
Here, $\overline{\sigma}$ denotes the spin species with opposite orientation to $\sigma$.
The key alignment property is contained in $\hat P_l$ which is designed to make the process conditional on a surrounding magnetization in such a way that spin-changing processes for a particle are suppressed or enhanced when the spin of the particle does or doesn't align with the neighborhood, respectively.
This can be achieved for various variants of $\hat P_l$ (see Supplementary Material Sec.~II).
In the following, we use $\hat P_l= \mathrm{exp}(-K/(2r)\hat{m}_l\sum_{|j|=1}^{r}\hat{m}_{l+j})$ with $K$ denoting the alignment parameter and $r$ defining the interaction radius, which we choose as $r=4$ in the following.
Here, $\hat m_l=\hat n_{\uparrow l} - \hat n_{\downarrow l}=\hat{c}^{\dagger}_{\uparrow l}\hat{c}_{\uparrow l}-\hat{c}^{\dagger}_{\downarrow l}\hat{c}_{\downarrow l}$ measures the local magnetization.
In analogy to the Vicsek model~\cite{Vicsek_1995}, where each point-agent aligns its velocity direction to the average direction of motion of its neighborhood,  our quantum particles align according to their surrounding magnetization.

The Hamiltonian contributes a quantum-coherent real-time evolution through a local spin-flip term 

$\hat H=-g\sum_{l=1}^{L}\big(\hat{c}_{l,\uparrow}^{\dagger}\hat{c}_{l,\downarrow}+h.c\big)$ with $g$ denoting the quantum amplitude.
This process provides an inherent (quantum) noise source that counteracts the flock formation.
Specifically, the Hamiltonian dynamics randomizes the spin degree of freedom of the particles overcoming at some strength of $g$ the alignment tendency from the dissipative dynamics.
Importantly, we will find that this Hamiltonian contribution also qualitatively changes the character of the flock motion inducing a long-distance quantum coherence absent in the classical case.

A key property of the considered model is a $\mathbb{Z}_2$-symmetry by flipping the spins of all particles upon simultaneously reversing the particles' direction of motion, which is the polar symmetry that will be spontaneously broken in the quantum flock.
Due to the link between the species type $\sigma$ and the direction of motion, the collective motion of particles is detectable through the magnetization $\hat M = \sum_{l=1}^L \hat m_l$ as the order parameter.

\begin{figure} 
\begin{center}
\includegraphics[width=\columnwidth]{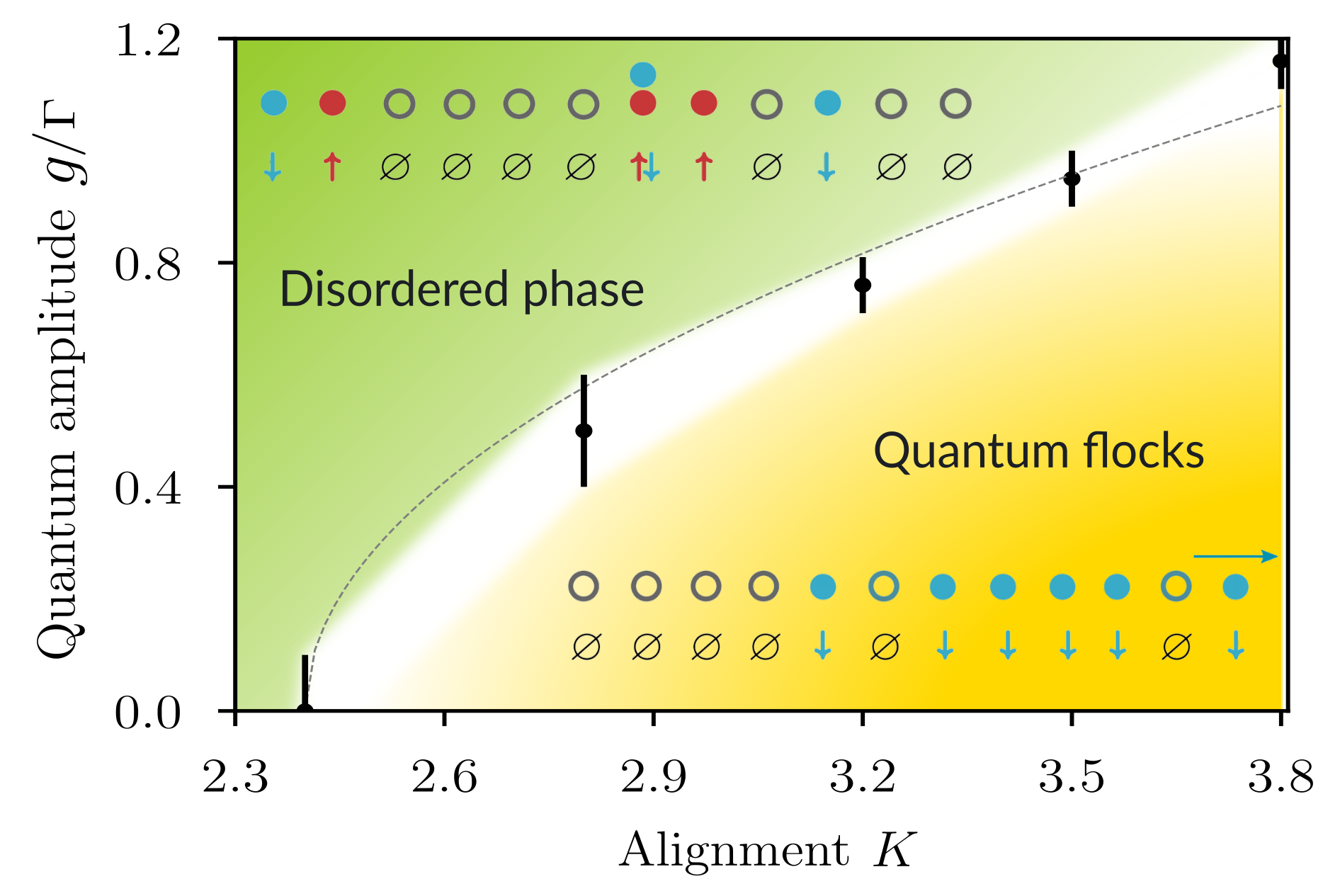}
\caption{\textbf{Phase diagram.} 
Numerically obtained phase diagram with indicated error bars as a function of the alignment parameter $K$ and the quantum amplitude $g$ including representative many-body configurations. The numerical uncertainty for the estimate of the phase transition points is indicated by the white region separating the disordered from the quantum flocking phase. The analytical prediction from the coarse-grained theory at weak quantum amplitudes $g$ is included as a dashed line, which is consistent with the numerical results.
}
\label{order_quantum:fig}
\end{center}
\end{figure}

\textit{Long-range order and collective motion:--} We target the detection of long-range order by means of the Binder cumulant $U(t)=1- \langle \hat M^4(t) \rangle/(3\langle \hat M^2(t)\rangle^2)$ associated to the order parameter $\hat M$.
Here, $\langle \hat{\mathcal{O}}(t) \rangle = \mathrm{Tr}( \hat \rho(t) \hat{\mathcal{O}})$ denotes the time-dependent expectation value of the operator $\hat{\mathcal{O}}$.
A more direct measure and visualization of the quantum flock by means of snapshot measurements we introduce below.
In the thermodynamic limit, the Binder cumulant is $U=2/3$ for long-range ordered states whereas $U=0$ for disordered ones.
In Fig.~\ref{fig:aqim}B, we display numerically obtained data for the Binder cumulant $U(t)$ with qualitatively different behavior depending on the strength of the quantum amplitude $g$. 
For weak $g/\Gamma=0.2$, the Binder cumulant rises up to $U(t) \approx 2/3 $ for long times with only weak finite-size effects.
These results suggest long-range order and the realization of an active quantum flock experiencing collective motion. 
The behavior is different in the opposite case of large quantum amplitudes $g/\Gamma=0.8$ where the attained long-time value exhibits a considerable system-size dependence with $U \to 0$ upon increasing $L$, indicating a disordered phase.
We note that all results are obtained at a fixed particle density \( \nu = N/L = 1/2 \).

In Fig.~\ref{fig:aqim}C, we show the long-time value $\overline{U}$ of the Binder cumulant as a function of $g$, obtained from a time average in the interval $\Gamma t \in [40,70]$. 
We observe compelling evidence for a long-range ordered phase at weak quantum amplitudes $g/\Gamma \ll 1$.
For large quantum amplitudes instead, the tendency is clearly towards a disordered state with $\overline{U} \to 0$ upon increasing $L$.
We estimate the phase transition point by identifying the value of $g$ at which $\overline{U}$ crosses a threshold $2/3-\epsilon$ with $\epsilon=0.02$, as indicated also in Fig.~\ref{fig:aqim}C.
The corresponding estimated phase diagram is shown in Fig.~\ref{order_quantum:fig}. The dashed line in this figure represents the analytical prediction derived from the coarse-grained description of the dynamics in the weak-$g$ limit, presented in the Coarse-Grained Dynamics section.
Alternatively, the flocks can also be characterized by the currents generated through the symmetry breaking, as detailed in Sec. II of the Supplementary Material. In the Sec. III of Supplementary Material we also display further evidence for long-range order in the quantum flocking phase through the real-space magnetization correlation function.

\begin{figure*}
\centering
\includegraphics[width=\textwidth]{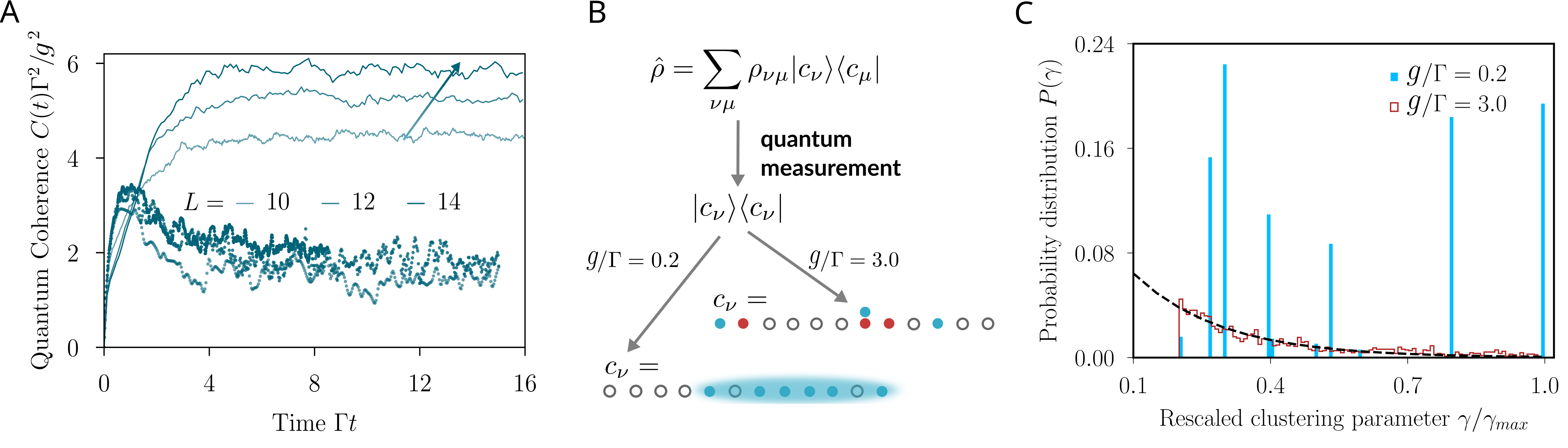}
\caption{\textbf{Quantum coherence and cluster formation.} \textbf{A} Dynamics of the total long-distance quantum coherence $C(t)$ comparing the flocking phase ($K=3.8$, solid lines) to the disordered phase ($K=0.5$, dotted lines) for $g/\Gamma=0.2$ and different system sizes $L$ calculated for $N_r=1000$ trajectories. \textbf{B} A quantum snapshot measurement on the full density matrix $\hat \rho$ yields many-body configurations $c_\nu$ with typical outcomes for the quantum flocking ($g/\Gamma=0.2$) and the disordered phase ($g/\Gamma=3.0$). \textbf{C} Cluster statistics $P(\gamma)$ comparing the quantum flocking ($g/\Gamma=0.2$) to the disordered phase ($g/\Gamma=3.0$) at  $K=3.8$ for $N_r=2000$ trajectories and $L=12$. Here, $\gamma_\text{max}=L^2/4$ denotes the maximally possible $\gamma$ value. 
}
\label{cluster_quantum:fig}
\end{figure*}

\textit{Quantum coherence:--} A key question remains, namely to what extent quantum flocks differ from those in the classical world.
Characterizing such differences is a challenging task especially in mixed states of quantum matter.
While we can identify quantum entanglement in the logarithmic negativity for the quantum flock (see Sec. IV of Supplementary Material) and thereby genuinely quantum behavior, in the following we aim to focus on the quantum coherence~\cite{Streltsov_2017}, which measures the amount of quantum superposition of a general mixed state.
We find that the quantum coherence displays the most striking qualitative difference between the quantum flocking and non-flocking phase, see Fig.~\ref{cluster_quantum:fig}A.
Concretely, we consider a total long-distance quantum coherence $C(t) = \sum_{l=1}^L C_l(t)$ with $C_l(t) = \sum_{\nu_l \not= \nu'_l} \big|\langle \nu_l | \hat \rho_{l}^{C}(t) | \nu'_l \rangle \big|$, where $\hat \rho_l^C(t)$ denotes the reduced density matrix of two lattice sites $l$ and $l+L/2$, i.e., at a maximal distance.
The states $|\nu_l\rangle=|\mathbf{n}_l,\mathbf{n}_{l+L/2}\rangle$ represent all the particle configurations with $\mathbf{n}_l=\varnothing,\uparrow,\downarrow,\uparrow\downarrow$.
Notice that for a vanishing quantum amplitude $g=0$ the dynamics reduces to a classical process and to classical flocking without quantum coherence and entanglement. Therefore, we will refer to a \textit{quantum} flock only for $g>0$.

In Fig.~\ref{cluster_quantum:fig}A, we show our numerical results for $C(t)$ at a fixed quantum amplitude $g/\Gamma=0.2$.
In the disordered phase for weak alignment $K=0.5$, $C(t)$ settles to a plateau independent of system size $L$.
In the flocking phase instead for large alignment $K=3.8$, $C(t)$ grows with increasing system size $L$ and time $t$.
Consequently, our numerical simulations suggest that the quantum flock exhibits distinct quantum properties through a long-distance quantum coherence.
The obtained results for $C(t)$ also have direct implications for the nature of the collective motion.
$C(t)$ is a measure on the strength of the off-diagonal matrix elements in $\rho_l^C(t)$, which involve correlation functions of the form $\langle \hat c_{l\uparrow}^\dag(t) \hat c_{l+L/2 \uparrow}(t) \rangle$ for instance.
This suggests that the quantum flocking state is not only characterized by quantum superposition but also by long-distance quantum-coherent motion.

\textit{Alignment and clustering:--} Having found evidence for intrinsic quantum effects, we aim in a next step to characterize the quantum flock on a microscopic level by means of a clustering analysis~\cite{Rodriguez_2014} extended to snapshot measurements.
Such snapshot measurements are obtained by performing a joint projective quantum measurement on each quantum degree of freedom providing as the outcome a single many-body configuration, as accessible on today's quantum simulation or computing platforms~\cite{Gross2017,Monroe2021,Browaeys2020,Blais2021}.
Typical snapshots are depicted in Fig.~\ref{cluster_quantum:fig}B pointing towards a fundamental difference between the flocking and disordered phases.
Without loss of generality, we focus on the $\sigma=\downarrow$-species with corresponding many-body configurations $\mathbf{n} = (n_{1\downarrow},n_{2\downarrow}, \dots, n_{L\downarrow})$ and $n_{l\downarrow}=0,1$.

Within the utilized clustering algorithm~\cite{Rodriguez_2014}, a clustering parameter $\gamma_l= \varrho_l \delta_l$ is associated to each lattice site $l$.
Here, $\varrho_l=n_{l \sigma} \sum_{|m-l|<d_c}   n_{m \sigma}$ denotes a coarse-grained local density and $\delta_l=\min_{m:\varrho_m>\varrho_l}(|l-m|)$ measures the distance to the next lattice site with higher density.
Large clusters can be identified through large values of $\gamma$ as they exhibit both a high local density $\varrho_l$ and a large distance $\delta_l$~\cite{Rodriguez_2014}.
In what follows, we choose $d_c=4$ for concreteness and for the site with the highest density we have $\delta_l=L/2$.

In Fig.~\ref{cluster_quantum:fig}C, we display the statistics of the clustering parameter $\gamma$ with $P(\gamma)$ obtained from a histogram.
$P(\gamma)$ displays a compelling difference between the two phases.
For large quantum amplitudes $g/\Gamma=3$, we find that $P(\gamma)$ exhibits a monotonically decaying behavior with a strongly suppressed probability for large clusters.
In the flocking phase, the picture is qualitatively different.
We find a sequence of isolated peaks at large $\gamma$ values which represent individual large clusters travelling throughout the system.

\textit{Coarse-grained dynamics:--} Based on the numerical evidence for a quantum flocking phase, it is a key next step to also develop an analytical understanding of our quantum flocking problem. 
We now present an analysis of a coarse-grained version of the Lindblad master equation, corroborating our numerical findings(see Supplementary Material, Sec. IX for details).
Let us first target a homogeneous solution for the magnetization, $m(t)=\langle \hat m_l (t) \rangle$ for all $l$.
We obtain as an exact result $ \dot m (t)  = g \langle \hat f \rangle
-2\Gamma_F\langle \hat P_l^\dagger\hat P_l \hat m_l\rangle$ (see End Matter).
Here, $\hat f(t) = -i (\langle \hat c_{l\uparrow}^\dag(t) \hat c_{l\downarrow}(t) \rangle - \langle \hat c_{l
\uparrow}(t) \hat c_{l\downarrow}^\dag(t) \rangle )$ captures the contribution from the  quantum dynamics.
In a next step we focus on large times $t$ with  $m=\lim_{t \to \infty}m(t)$.
Motivated by our numerical simulations suggesting a phase transition, we consider in analogy to a conventional Landau approach the limit of a small magnetization $m$, which allows us to perform expansions in powers of $m$.
By means of suitable mean-field factorizations we finally obtain (see End Matter):
\begin{equation}
    0= (K_c - K) m + \frac{4K}{K_c}\left(Kq -1 \right) m^3 + \mathcal{O}(m^5) \, .
    \label{eq:phi4}
\end{equation}
This result is reminiscent of the mean-field solution of the classical Ising model exhibiting a symmetry-broken phase for $K>K_c$ thereby confirming the existence of a quantum flocking phase.
%
Here, \( K_c = \frac{1+\Delta_g}{2\sigma_m^2} \), with \( \Delta_g = \frac{g^2}{2\Gamma^2} \) denoting the shift of the critical point induced by weak quantum amplitudes (\( g/\Gamma \ll 1 \)). This expression is consistent with the numerical results in Fig.~\ref{order_quantum:fig}, where the fitted critical points yield \(\sigma_m \approx 0.45\) and confirm the increase of \(K_c\) with larger \(g\).
In the derivation of Eq.~(\ref{eq:phi4}) we assume that higher moments of the magnetization can be expanded according to $\langle m^2\rangle = \langle m \rangle^2 + \sigma_m^2$ and $\langle m^3 \rangle = \langle m \rangle^3 + q\langle m \rangle$.
The above effective description is well justified for $r\gg1$, thereby providing a direct connection to the present coarse-grained theory in a well-controlled limit.
Importantly, the quantum flock requires $q > K_c^{-1}$ as the homogeneous solution becomes unstable otherwise (see End Matter).
Eq.~\eqref{eq:phi4} suggests a continuous phase transition for the quantum flocking problem.
This aligns with the properties of classical active matter systems in one dimension~\cite{czirok1999collective, o1999alternating}.

In view of the Mermin-Wagner-Hohenberg theorem, the finding of long-range order in one dimension might appear remarkable.
This, however, can be attributed solely to the active nature of the system and the breaking of  local detailed balance.
We further corroborate the existence of a long-range ordered state by means of a simulation of a classical analogue of the Lindblad master equation, where we also find evidence for a flocking phase for large system sizes.
Our coarse-grained theory also supports inhomogeneous solutions for $\Gamma_{\mathcal{A}}>\Gamma_\mathcal{M}$, where we observe the formation of traveling wave patterns (see Supplementary Material Sec.~V).
%
%

\textit{Discussion:--} In this work, we have introduced the concept of active quantum matter.
We have formulated a model for active quantum particles giving rise to quantum flocks with distinct quantum features by means of a long-distance quantum coherence.
It is a natural question to which extent the introduced quantum flocks might also be accessible experimentally.
In Supplementary Material Sec.~VIII, we present experimental schemes to realize in systems of Rydberg atoms the building blocks of the individual processes appearing in the considered Lindblad master equation.
%
%

%
For the future it will be central to further explore the details of the considered model such as to study its density dependence, which is an important control parameter for the classical flocking problem~\cite{Vicsek_1995,
bertin2009hydrodynamic,
weber2012nucleation, weber2013role,Solon_2013,Solon_2015}.
It will be further important to also target more specifically the quantum flocking transition.
On the basis of a coarse-grained description we have found evidence for both first-order and continuous transitions depending on the parameter regime.
For a numerical approach it would be necessary to explore other advanced numerical methods such as tensor networks~\cite{Orus2019} or neural quantum states~\cite{Carleo2017,Schmitt2020}, as we are operating already at the frontier of what is possible via exact diagonalization.
This would also allow us to explore the connection to other long-distance quantum-coherent motion such as in superfluids.

The present work paves the way to explore further active quantum matter systems, for instance by drawing inspiration from the classical side where also various other interesting nonequilibrium phases have been discovered such as motility-induced phase separation~\cite{cates2015motility}, active nematics~\cite{giomi2015geometry}, or intermittent collective motion~\cite{huepe2004intermittency, cavagna2013boundary, GomezNava2022}. 
It would also be a natural and promising next step to move towards higher dimensions.
Of particular interest would be that higher dimensions might also enable the spontaneous breaking of more complex symmetries than $\mathbb{Z}_2$.
Overall, we expect that our work will pave the way to yet unexplored nonequilibrium phases of quantum matter with intriguing properties.

~\\
\textit{\mbox{Data availability\hspace{0.05mm}\llap{Data availability}} }
~\\
\noindent
The data displayed in the figures is available on dx.doi.org/10.5281/zenodo.8208685.

~\\
\noindent {\it Acknowledgements:-} We acknowledge valuable discussions with Ricard Alert, Rainer Blatt, Alexander Eisfeld, Juan Garrahan, Michael Knap, Igor Lesanovsky, Frank Pollmann, Achim Rosch, and Johannes Zeiher.
This project has received funding from the European Research Council (ERC) under the European Union’s Horizon 2020 research and innovation programme (grant agreement No. 853443), and M. H. further acknowledges support by the Deutsche Forschungsgemeinschaft via the Gottfried Wilhelm Leibniz Prize program.
Parts of the numerical simulations were performed at the Max Planck Computing and Data Facility in Garching.

\bibliography{literature}

\section*{End Matter}
~\\
\noindent
{\it Numerical solution of the Lindblad master equation:-} We solve the Lindblad master equation via exact diagonalization through a mapping to a stochastic Schr\"odinger equation.
We achieve up to $L=14$ lattice sites, which on the level of the Lindblad equation corresponds to solve $10^{12}$ coupled first-order differential equations. 
In our simulations, we choose for convenience $\Gamma_\mathcal{M}=\Gamma_\mathcal{A}=\Gamma$, and we consider initial conditions $|\psi_0\rangle$ with a vanishing magnetization and short-range correlations: $|\psi_0\rangle = \otimes_{l=1}^{N} |\psi_0\rangle_l \otimes_{l=N+1}^L \ket{0}_l$, where $|\psi_0\rangle_l = 2^{-1/2}(\ket{\uparrow}_l + \ket{\downarrow}_l)$ and $|0\rangle_l$ denotes an empty lattice site.
We verified that the properties of the steady state do not depend on the choice of the initial condition (see Supplementary Material Sec. VII).
The initial condition sets the density $\nu=N/L=1/2$ of particles in the system.

We solve numerically the Lindblad master equation in Eq.~(\ref{eq:Lindblad}) as a piece-wise deterministic process.
Instead of calculating the full dynamics of the density matrix $\hat{\rho}(t)$ we sample pure-state trajectories $\ket{\psi_t(z)}$ in Hilbert space according to a probability distribution such that we recover $\hat{\rho}_t$ as an average over the individual trajectories~\cite{schaller2014open}, i.e.,
\begin{equation}
\rho_t={\cal M}\{\ket{\psi_t(z)}\bra{\psi_t(z)}\}~,
\end{equation}	
where $z$ refers to a suitably chosen stochastic process. The evolution of the system can then be effectively modeled by~\cite{schaller2014open}
\begin{equation}\label{eq:SDE}
\begin{split}
\ket{d\psi}&=\hat{H}_{eff}\ket{\psi}dt+\sum_{l,\sigma}\Big(\frac{\hat{\mathcal{M}}_{l,\sigma}\ket{\psi}}{\sqrt{\braket{\psi|\hat{\mathcal{M}}_{l,\sigma}^{\dagger}\hat{\mathcal{M}}_{l,\sigma}|\psi}}}-\ket{\psi}\Big)dN_{l,\sigma}^\mathcal{M}\\&+\sum_{l,\sigma}\Big(\frac{\hat{\mathcal{A}}_{l,\sigma}\ket{\psi}}{\sqrt{\braket{\psi|\hat{\mathcal{A}}_{l,\sigma}^{\dagger}\hat{\mathcal{A}}_{l,\sigma}|\psi}}}-\ket{\psi}\Big)dN_{l,\sigma}^\mathcal{A}~,
\end{split}
\end{equation}
and therefore by a non-linear stochastic Schr\"odinger equation. The effective non-Hermitian Hamiltonian reads as~\cite{schaller2014open}
\begin{equation}
\begin{split}
&\hat{H}_{eff}=-iH-\frac{\Gamma_{\mathcal{M}}}{2}\sum_{l,\sigma}\hat{\mathcal{M}}_{l,\sigma}^{\dagger}\hat{\mathcal{M}}_{l,\sigma}-\frac{\Gamma_{\mathcal{A}}}{2}\sum_{l,\sigma}\hat{\mathcal{A}}_{l,\sigma}^{\dagger}\hat{\mathcal{A}}_{l,\sigma}\\&+\frac{\Gamma_{L}}{2}\sum_{l,\sigma}\braket{\psi|\hat{\mathcal{M}}_{l,\sigma}^{\dagger}\hat{\mathcal{M}}_{l,\sigma}|\psi}+\frac{\Gamma_{\mathcal{A}}}{2}\sum_{l,\sigma}\braket{\psi|\hat{\mathcal{A}}_{l,\sigma}^{\dagger}\hat{\mathcal{A}}_{l,\sigma}|\psi}~.
\end{split}
\end{equation} 
The Poisson increments $dN^X_{l,\sigma}$ satisfy
\begin{equation}\label{Poisson_increments}
dN^X_{\alpha}dN^X_{\beta}=\delta_{\alpha\beta}dN^X_{\alpha},~~~~M(dN^X_{\alpha})=\Gamma_{X}\braket{\psi|\hat{X}_{\alpha}^{\dagger}\hat{X}_{\alpha}|\psi}~,
\end{equation}
where $M(X)$ indicates the classical average over the trajectory ensemble, and $dN^X_{\alpha}$ can be $\{0,1\}$.
Equation~(\ref{Poisson_increments}) implies that we have at most one single jump at each time step $t$ occurring with probability $P_{\alpha}=\sum_x\Gamma_{X}\braket{\psi|\hat{X}_{\alpha}^{\dagger}\hat{X}_{\alpha}|\psi}dt$.
On a general level, the stochastic differential equation in Eq.~(\ref{eq:SDE}) represents a combination of a deterministic evolution and stochastic quantum jumps.
The stochastic part we solve as follows.
First, at a given time $t$, we evaluate for the next time step the total jump probability during the interval $[t,t+\Delta t]$ with $\Delta t$ a small time interval.
Based on that probability we randomly decide for the occurrence of a jump.
In case a jump is supposed to take place, we further randomly select the type of the jump, which is then finally executed.
In the opposite case of no jump, we replace $dN_{\alpha}=0$ for all $\alpha$ in Eq.~(\ref{eq:SDE}) and solve the resulting deterministic non-linear differential equation to determine $\ket{\psi(t+\Delta t)}$ at the next step.
This procedure is then iterated over time in order to obtain a full trajectory of the quantum many-body state $\ket{\psi_t(z)}$.
In the end we average over $N_r$ such trajectories to calculate expectation values of observables.

{\it Coarse-grained dynamics:-} Here, we provide details on the derivation of the coarse-grained 
equations of motion obtained from the Lindblad master equation.
The  local occupation numbers $\hat{n}_{l, \sigma} = \hat c_{l\sigma}^\dag \hat c_{l\sigma}$ obey the following general expression:
\begin{align}
\begin{split}
\frac{d}{d t}  & \langle \hat{n}_{l, \sigma} \rangle =  
\sigma g \langle \hat{f}_l\rangle + \Gamma_\mathcal{A} 
\langle \hat{P}_l^\dagger \hat{P}_l 
 \left(\hat{n}_{l\overline{\sigma}} - \hat{n}_{l\sigma} \right)\rangle+\\
&+ \Gamma_\mathcal{M} \left[ \langle \hat{n}_{l+\sigma, \sigma} \left( 1-\hat{n}_{l, \sigma}\right)\rangle 
-\langle \left( 1-\hat{n}_{l -\sigma, \sigma}\right)  \hat{n}_{l,\sigma} \rangle
\right]
\end{split}
\label{eq:hydro_n}
\end{align}
with $\sigma = \up,\down$ and $\overline{\sigma}$ being the complement 
of $\sigma$. In the spatial index we also use 
$\up/\down = \pm 1$ and 
$\hat{f}_l = i(\hat{c}_{l\up}^\dagger \hat{c}_{l\down} -\hat{c}_{l\up} \hat{c}_{l\down}^\dagger)$ is the local spin flip current operator.

Interestingly, Eq.~(\ref{eq:hydro_n}) reproduces the classical TASEP dynamics in the limit $g,\Gamma_F \to0$~\cite{Schue03}. From Eq.~(\ref{eq:hydro_n}) we derive the equations of motion for the local densities $\hat{\rho}_l = (\hat{n}_{l\up}+\hat{n}_{l\down})/2$ and magnetizations $\hat{m}_l = (\hat{n}_{l\up}-\hat{n}_{l\down})/2$, viz.,
\begin{align}
 \frac{d}{d t} \langle\hat \rho_l\rangle&= \Gamma_\mathcal{M}
 \big[ 
 \langle (\frac{1}{2}-\hat\rho_l)\delta\hat m_{l}\rangle
 -\langle \hat m_l\delta \hat \rho_{l}\rangle 
 +\frac{\langle \delta^2 \hat{\rho}_l \rangle}{2} 
 \big]
 \label{eq:hydro_rho}
\\[.5cm]
\label{eq:ml}
\begin{split}
 \frac{d}{d t} \langle \hat m_l\rangle&=  
 g \langle \hat f_l \rangle
 -2\Gamma_\mathcal{A}\langle \hat P_l^\dagger\hat P_l \hat m_l\rangle\\
  &-\Gamma_\mathcal{M} \big[\langle
 (\hat\rho_l-\frac{1}{2})\delta \hat\rho_{l}\rangle +\langle\hat m_l\delta\hat m_{l}\rangle
 -\frac{\langle \delta^2\hat m_l\rangle }{2}  \big]
 \end{split}
\\[.5cm]
  \label{eq:fl}
  \begin{split}
 \frac{d}{dt}\langle\hat f_l\rangle &=  
  -4g\langle\hat m_l\rangle
 - \frac{\Gamma_\mathcal{A}}{2}\langle
\{\hat  P_l^\dagger \hat P_l, \hat f_l\}\rangle\\
&\hspace{3cm}-\Gamma_\mathcal{M} \langle(1+\delta \hat m_{l})\hat f_l\rangle
\end{split}
\end{align}
with the finite differences $\delta \mathcal{O}_l =\mathcal{O}_{l+1}-\mathcal{O}_{l-1}$ and $\delta^2 \mathcal{O}_l =\mathcal{O}_{l+1} + \mathcal{O}_{l-1} - 2\mathcal{O}_l$. In the remainder, we assume $[\hat{P}_l^\dagger\hat{P},\hat{f}_l] = 0$ which is 
true, e.g., when $\hat{P}_l^\dagger\hat{P}_l$ is a function of the local magnetizations.

{\it Homogeneous mean-field solution.}
First, we study homogeneous solutions of Eqs.~(\ref{eq:hydro_rho})-(\ref{eq:fl}) where $m(t)=\langle \hat m_l(t) \rangle$ and $\rho(t) = \langle \hat \rho_l(t) \rangle$ for all $l$.
Upon factorizing the term proportional to $\Gamma_{\cal M}$ in a mean-field way all correlations, Eq.~(\ref{eq:ml}) reads:
\begin{equation}
	\frac{d}{dt} m(t) = g f(t) - 2\Gamma_\mathcal{A} \langle \hat P_l^\dag(t) \hat P_l(t) \hat m_l(t) \rangle
\end{equation}
Here, we address the nonlinearity $\langle \hat P_l^\dag(t) \hat P_l(t) \hat m_l(t) \rangle$ by using the explicit form of the alignment operator
 $\hat{P}_l^\dagger \hat{P}_l = \exp \left( -2K\hat{m}_l \hat{M}^{l}\right)$,
with  $\hat M^{l}=L^{-1}\sum_{j\not=l} \hat m_{j}$. 
This corresponds to the alignment operator in the main text for $r=L/2$.
In analogy to a conventional Landau description of the phase transition, we capture the onset of order using a Taylor-expansion in powers of the magnetization:
\begin{align}
\label{eq:PP}
  \langle \hat m_l \hat P_l^\dagger \hat P_l \rangle \simeq m - 2K \langle \hat m_l^2 \hat M^{l} \rangle  +2K^2 \langle \hat m_l^3 \hat M^l \hat M^l \rangle \, .
\end{align}
Next, we address the factorization of the 
remaining higher order correlations.
To this end, we consider all local magnetizations as independently fluctuating, uncorrelated quantities.
As $\hat M^l$ is a mean magnetization, we expect its fluctuations to be subleading 
for large system sizes and thus neglect them. We then find:
\begin{align}
	\label{eq:PP_homogeneous}
	\langle \hat m_l P_l^\dagger P_l \rangle \simeq m - 2K \langle \hat m_l^2 \rangle m  +2K^2 \langle \hat m_l^3 \rangle m^2 
\end{align}
where we have used $\langle \hat M^l \rangle = m$ for the targeted homogeneous solution.
Thus, the higher moments of the local magnetization remain to be considered. We expand these in terms of lower-order moments according to

 $\langle \hat m_l^2 \rangle = m^2 + \sigma^2$
 and
 $\langle \hat m_l^3 \rangle = m^3 + q m$,
with $\sigma^2$ and $q$ some constants, whose exact values would have to be derived from 
microscopic considerations.
Consequently, for $g=0$, we obtain for the stationary steady state solution:
\begin{align}
\frac{dm}{dt} = 0 = {m}^4 + \left[ q -\frac{1}{K}\right] m^2 +   \frac{1}{2K^2} - \frac{\sigma^2}{K}  .
\end{align}
For the onset of the flocking phase with $|m|\ll 1$, we can neglect the contributions of the order $m^4$ yielding:
\begin{align}
m^2 \simeq \frac{1}{2K_c^2}\frac{ K/K_c - 1}{ q - 1/K_c}
\label{eq:m_homo}
\end{align}
with the critical value $K_c = 1/(2\sigma^2)$.
This equation exhibits solutions along two distinct branches.
One for $K>K_c$ with $q>1/K_c$, and the other for $K<K_c$ with $q<1/K_c$.
As we show now, only one of these is stable.
We consider a weak, and still homogeneous, time-dependent deviation $u(t)$ on top of the homogeneous solution, i.e., 
$ m = m_0 + u(t)$.
We find for the dynamics of the deviation to leading order:
\begin{align}
    \frac{du}{dt} = -4\Gamma_\mathcal{A}\left(\frac{K}{K_c}-1\right)u
\end{align}
which clearly indicates that only the solution with $K>K_c$ and $q>1/K_c$ is stable.

Lastly, we consider the influence of the quantum dynamics of amplitude $g$ on the homogeneous solution~(\ref{eq:m_homo}).
The quantum dynamics couples Eq.~(\ref{eq:ml}) and Eq.~(\ref{eq:fl}). In the stationary state we find the homogeneous
transverse current $f  \simeq - 4gm/(\Gamma_\mathcal{M} + \Gamma_\mathcal{A})$.
Hence, the homogeneous solution is altered as follows
\begin{align}
    m^2 \simeq \frac{1}{2K_c^2}\frac{K/K_c - 1 - \Delta_g}{q-1/K_c}
\end{align}
with $\Delta_g = 4g^2/(\Gamma_\mathcal{A}(\Gamma_\mathcal{M}+\Gamma_\mathcal{A}))$. In particular we find that the quantum dynamics shifts the critical point as $K_c(g) \simeq (1+\Delta_g) K_c(g=0)$.
Since $\Delta_g >0$, larger coupling strengths are required to induce order, which is consistent with the exact numerical data for the phase diagram.
In the Supplementary Material we further discuss the inhomogeneous solutions provided by the coarse-grained description yielding also traveling wave patterns. 
\begin{figure}[t]
\centering
\includegraphics[width=\columnwidth]{Fig4_maintext.pdf}
\caption{
\textbf{Classical analogue of the quantum flocking problem.} Dynamics of magnetization fluctuations \(M^2(t)\) for the alignment parameter \(K=0.5\) in \textbf{A} and \(K=3.5\) in \textbf{B}.
}
\label{fig:order_t-classic_main}
\end{figure}
{\it Classical limit:-} Our flocking model discussed in the main text exhibits a classical limit for $g \to 0$.
In this limit, the coherent spin-flip dynamics generated by the Hamiltonian is absent, leaving behind purely stochastic processes: biased hopping, alignment interactions, and effective noise.
While the quantum model features coherent and dissipative competition, the classical counterpart consists solely of probabilistic jump processes, making it also computationally simple.

The behavior of the classical model concerning flocking is similar to the quantum case. In Fig.~\ref{fig:order_t-classic_main}, we simulate the evolution of the order parameter fluctuations $M^2$ for two values of the alignment strength, $K=0.5$ and $K=3.5$, and various system sizes $L$.
For weak alignment ($K=0.5$), the long-time value of $M^2$ decreases with increasing $L$, indicating the absence of long-range order.
In contrast, for strong alignment ($K=3.5$), $M^2$ remains finite as $L$ increases, signaling the emergence of a stable, symmetry-broken phase with global magnetization — the classical case of flocking.
These results demonstrate that also in the absence of coherent dynamics, the essential ingredients of flocking — directional motion and alignment — are sufficient to induce collective behavior. Additional numerical data are provided in the Supplementary Material.

\renewcommand{\changed}[1]{#1}
\input{supplement_arxiv.tex}

\end{document}

%% file: supplement_arxiv.tex
\clearpage

\onecolumngrid

\begin{center}
    {\bf \large Supplemental Materials: Active quantum flocks}
\end{center}


\setcounter{equation}{0}
\setcounter{figure}{0}
\setcounter{table}{0}
\setcounter{section}{0}
\setcounter{page}{1}
\makeatletter
\renewcommand{\theequation}{S\arabic{equation}}
\renewcommand{\thefigure}{S\arabic{figure}}


\setcounter{section}{0}
\setcounter{page}{1}

\renewcommand{\figurename}{Fig.}
\setcounter{figure}{0}

\section{Violation of Kolmogorov's criterion}\label{Kolmogorov:app}

The Kolmogorov criterion provides a sufficient and necessary condition for a system to satisfy detailed balance.
In order to show that a system breaks detailed balance, it is sufficient to identify a multi-stage process $1\to 2 \to 3 \to ... \to M$, where the last state $M=1$ is identical to the first one in such a way that the total transition rate of the forward process:
\begin{equation}
    \Gamma_F = \Gamma_{1\to 2} \Gamma_{2 \to 3} \cdots \Gamma_{M-1\to 1}
\end{equation}
and the backward process
\begin{equation}
    \Gamma_B = \Gamma_{1 \to M-1} \cdots \Gamma_{3 \to 2} \Gamma_{2 \to 1}
\end{equation}
are not identical, i.e., $\Gamma_F \not = \Gamma_B$.

In the following, we will identify such a process in the quantum model.
For that purpose consider in a first step a slightly generalized model, as compared to the main text, where the hopping of the particles isn't fully unidirectional.
Concretely, we choose the rates for jumps of $\uparrow$-particles to be $\Gamma(1+\epsilon)/2$ to the left and $\Gamma(1-\epsilon)/2$ to the right.
Analogously, we define the rates for jumps of the $\downarrow$-particles to be $\Gamma(1-\epsilon)/2$ to the left and $\Gamma(1+\epsilon)/2$ to the right.
For $\epsilon=1$ this reduces to the dynamics we have for the quantum flocking problem in the main text.
At $\epsilon=0$ the jump rates are fully symmetric.
Further, let's choose the transition rates for the alignment process as in the original model, but with $r=1$ for simplicity (without loss of generality).

\begin{figure}[h]
    \centering
    \includegraphics[width=0.7\columnwidth]{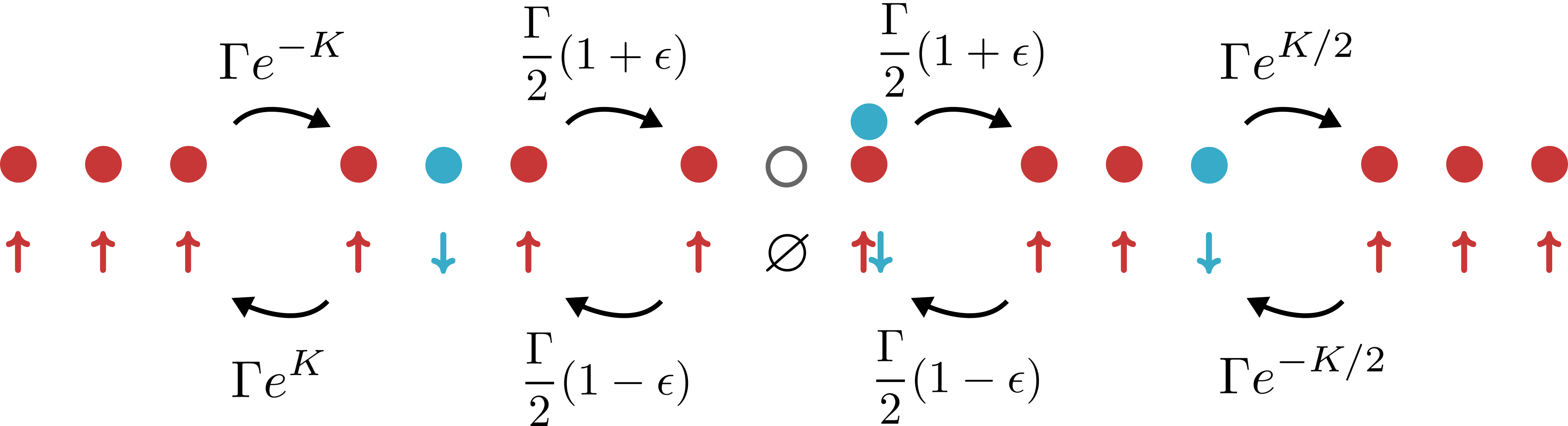}
    \caption{Violation of Kolmogorov's criterion for a sequence of transitions starting from a configuration of three spin-$\uparrow$ particles, assuming that the neighboring lattice sites are empty. The transition rates for the individual processes are included.}
    \label{fig:kolmogorov}
\end{figure}

Following a similar argument as in Refs.~[S1,S2] for classical active Ising models, consider now a configuration of three neighboring $\uparrow$-particles and a sequence of transitions as illustrated in Fig.~\ref{fig:kolmogorov}.
For this sequence we find $\Gamma_F=\Gamma^4 e^{-K/2}(1+\epsilon)^2/4$ and $\Gamma_B=\Gamma^4 e^{K/2}(1-\epsilon)^2/4$.
Clearly, we observe that $\Gamma_F \not= \Gamma_B$, in particular in the case $\epsilon=1$ discussed in the main text.
Therefore, Kolmogorov's criterion is violated and the system breaks local detailed balance.
Notice that this holds even in the fully symmetric case $\epsilon=0$ where we have that $\Gamma_F \not= \Gamma_B$ as long as $K\not=0$.
The presented argument doesn't involve the Hamiltonian part of the Lindblad master equation in the main text, but is rather solely based on the dissipative contributions to the dynamics.
Consequently, the breaking of local detailed balance is caused by the coupling to the environment.

%
\begin{figure*}
\centering
\includegraphics[width=\textwidth]{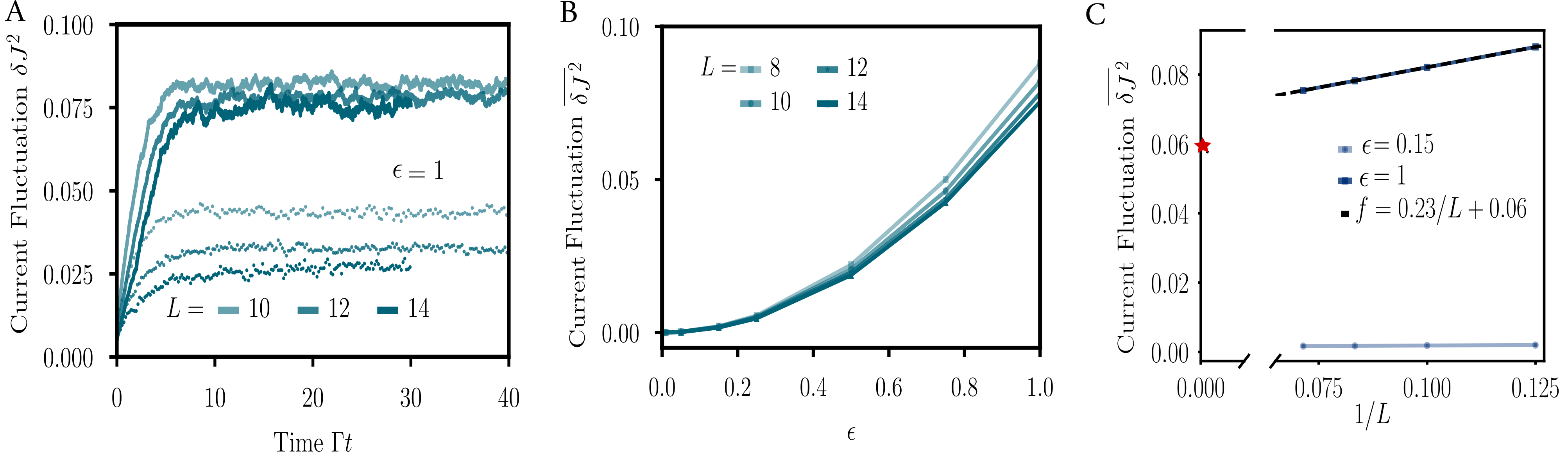}
\caption{\textbf{Current fluctuations.} \textbf{A} Current fluctuations over time for $\epsilon=1$. Here, dots represent cases where $g=3.0$, and lines indicate cases where $g=0.2$. \textbf{B} Variation of current fluctuations with $\epsilon$. We fix $g=0.2$ and $K=3.8$. \textbf{C} Size-dependent behavior of current fluctuations. This graph examines how current squared changes with different system sizes.
}
\label{Current_fluc:fig}
\end{figure*}
\changed{
\section{Current and Current Fluctuations}
On a general level flocks are associated with a symmetry breaking, that leads to directed motion.
For the active quantum flocking model studied in the main text the direction of motion is inherently linked to the particle species as an internal degree of freedom.
This is why in the main text the magnetization is chosen as the order parameter for the quantum flocking problem.
Alternatively, the flocks can also be characterized by the currents, which are generated through the symmetry breaking.
It will be the goal of this section to show that the quantum flocking phase described in the main text, and characterized by long-range magnetic order, also is associated with nonvanishing currents.
\begin{equation}
\begin{split}
    \hat{J}_l= & \frac{1+\epsilon}{2}\bigg[(1-\hat{n}_{l+1\downarrow})\hat{n}_{l\downarrow}-(1-\hat{n}_{l\uparrow})\hat{n}_{l+1\uparrow}\bigg] \\
    & + \frac{1-\epsilon}{2}\bigg[(1-\hat{n}_{l+1\uparrow})\hat{n}_{l\uparrow}-(1-\hat{n}_{l\downarrow})\hat{n}_{l+1\downarrow}\bigg]
\end{split}
\end{equation}
Notice that this current operator actually refers to a more general situation as it is considered also in Sec.~\ref{Kolmogorov:app} where the jumping rates for $\uparrow$-particles is set to $\Gamma(1+\epsilon)/2$ for leftward moves and $\Gamma(1-\epsilon)/2$ for rightward moves.
Likewise, the jumping rates for the $\downarrow$-particles are $\Gamma(1-\epsilon)/2$ to the left and $\Gamma(1+\epsilon)/2$ to the right.
This allows us to tune the degree of the breaking of detailed balance of the hopping contribution from $\epsilon=0$, where for each particle the motion is fully symmetric, to $\epsilon=1$ considered in the main text, where the motion is totally asymmetric.}

\changed{
Let us start by noting that the net current across the entire system is vanishing, even in the flocking phase due to symmetry in that current \(J=1/L\sum_l \langle \hat{J}_l \rangle = 0\). 
Physically speaking, in the flocking phase, each trajectory randomly chooses a preferred direction of motion, either to the left or to the right, resulting in an overall balance with no net flow.
Still, the flock can be characterized through currents in terms of the current fluctuations
\begin{equation}
    \delta J^2=\langle\hat{J}^2\rangle-\langle\hat{J}\rangle^2 \, ,
\end{equation}
which we will explore in the following.
For the flocking phase one expects $\delta J >0$ whereas $\delta  J \to 0$ for the disordered phase in the thermodynamic limit.
}

\changed{
In Fig.~\ref{Current_fluc:fig}A, we present the temporal evolution $\delta J^2(t)$ at $\epsilon=1$ and $K=3.8$ utilized also in the main text for weak ($g/\Gamma=0.2$) and strong ($g/\Gamma=3$) quantum amplitudes.
We observe that in the disordered phase ($g/\Gamma=3$), the current fluctuations decrease with increasing system size.
In contrast, within the quantum flock phase, $\delta J^2$ remain largely stable regardless of system size.
We further examine in more detail the long-term average $\overline{\delta J^2}$ of $ \delta J^2(t) $ as a function of the asymmetry parameter $\epsilon$, as shown in Fig.~\ref{Current_fluc:fig}B.
When $\epsilon$ is small, close to the fully symmetric case, the long-term average $\overline{\delta J^2}$ tends toward zero.
However, for larger $\epsilon$ also $\overline {\delta J^2}$ increases with weak finite-size dependence.
In order to quantify the system-size dependence, we further display $\overline{\delta J^2}$ as a function of $1/L$ in Fig.~\ref{Current_fluc:fig}C for $\epsilon=0.15$ and $\epsilon=1$.
In particular, for $\epsilon=1$ the current appears to exhibit a linear dependence on $1/L$.
Assuming that this behavior continues for even larger system sizes, an extrapolation through a linear fit yields nonvanishing current fluctuations $\overline{\delta J^2} \approx 0.06$ for $L\to \infty$ further supporting the evidence for a quantum flocking phase in the main text.
}
\changed{
\section{Real-space correlations in quantum flocks}
To further characterize the active quantum flock presented in the main text we additionally study in this section real-space profile of the magnetization correlations $m_d$ in the steady state:
\begin{equation}
  m_{d} = L^{-1}\sum_l \langle \hat m_l \hat m_{l+d} \rangle  
\end{equation}
as a function of distance $d$, see Fig.\ref{fig:Log_neg_corr}A-B.
In Fig.~\ref{fig:Log_neg_corr} we include $m_d$ for the flocking phase at $g/\Gamma=0.2$ (Fig.~\ref{fig:Log_neg_corr}A) and the disordered phase at $g/\Gamma=3$ (Fig.~\ref{fig:Log_neg_corr}B).
As one can see for largest accessible system sizes, in the flocking phase these magnetization correlations settle to a constant nonzero plateau at large distances $d$ consistent with long-range order, whereas they decay in the disordered phase.
}
\changed{
\begin{figure*}
    \centering
    \includegraphics[width=0.25\textwidth]{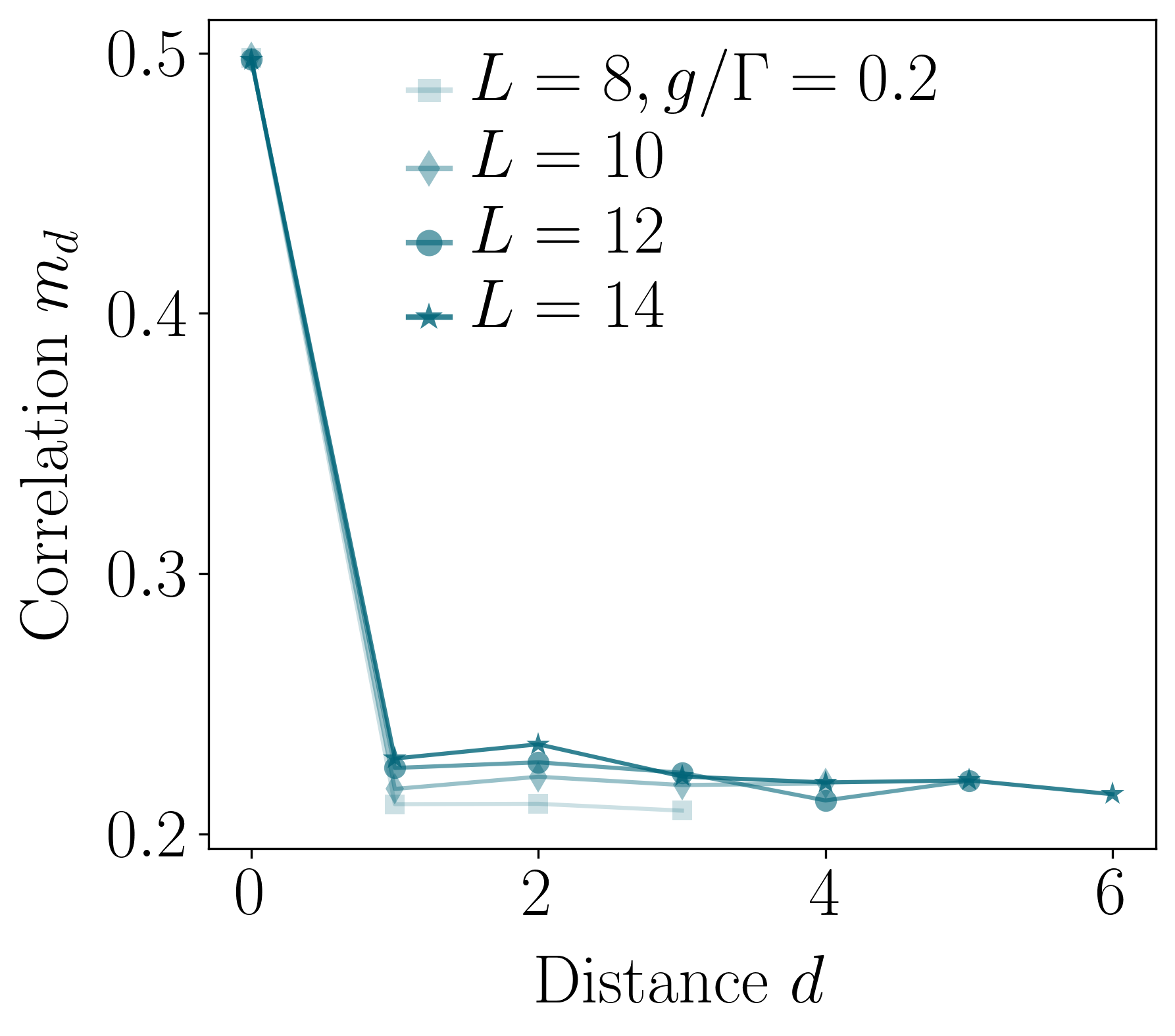}\put(-120,115){A}
     \includegraphics[width=0.228\textwidth]{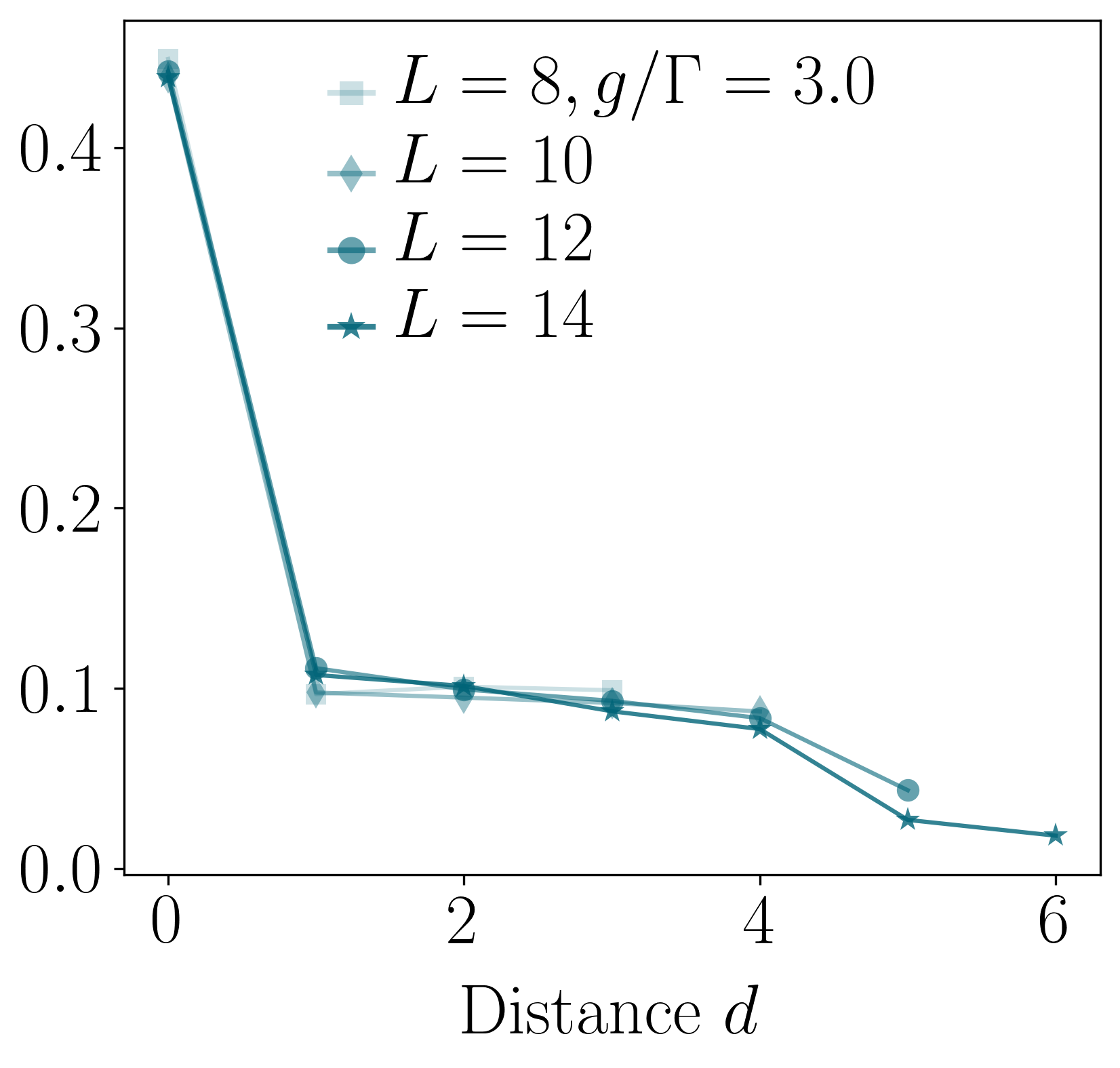}\put(-120,115){B}
    \includegraphics[width=0.28\columnwidth]{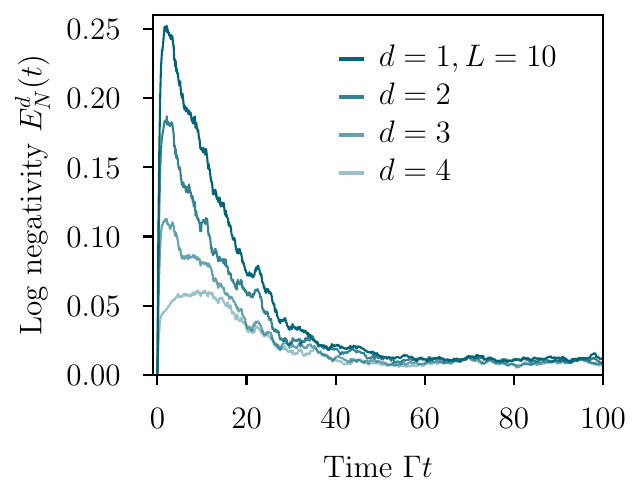}\put(-120,115){C}
    \caption{\textbf{Real space correlations and logarithmic Negativity}. Magnetization correlations as a function of distance $d$ for \textbf{A} $g/\Gamma=0.2$ \textbf{B} $g/\Gamma=3.0$. The remaining parameters are chosen as in the main text. \textbf{C} The dynamics of logarithmic negativity \(E_N^d(t)\) in the flocking phase with \(g=0.8\). The results show $E_N^d(t)$ between \(\uparrow\) and \(\downarrow\) particles on two lattice sites separated by varying distances \(d\). The alignment strength is \(K=3.8\) for all plots.
}
    \label{fig:Log_neg_corr}
\end{figure*}
\section{Quantum entanglement: Logarithmic Negativity}
As already emphasized in the main text, it is a key question to which extent the flocks we observe in our quantum model actually carry unique quantum features.
In the main text, we focus on quantifying quantum signatures through a quantum coherence.
In this section of the Supplementary Material we further aim to substantiate quantum properties by studying quantum entanglement.
}

\changed{
When it comes to entanglement quantification in mixed states, the
logarithmic negativity provides a particularly valuable and computable tool, as many traditional measures, such as entanglement entropy, are not applicable beyond pure states.
The logarithmic negativity $E_N$ is defined through the partial transpose $T_A$ operation of a density matrix $\rho$ of a quantum system.
It can be computed as follows:
\begin{equation}
    E_N = \log_2 \|\rho^{T_A}\|_1,
\end{equation}
where $\rho^{T_A}$ represents the partial transpose of $\rho$ with respect to subsystem $A$, and $\|\cdot\|_1$ denotes the trace norm, which is the sum of all singular values. 
}

\changed{
In the following we will focus on characterizing inter-species quantum entanglement, i.e., the entanglement between the $\uparrow$ and $\downarrow$ particles..
Concretely, we consider the reduced density matrix $\rho_{l,l+d}$ of two lattice sites $l$ and $l+d$ separated by a distance $d$, as for the quantum coherence in the main text.
We calculate the quantum entanglement on these two lattice sites between the $\uparrow$ and $\downarrow$ particles by performing a partial transpose $T_\uparrow$ (or equivalently $T_\downarrow$) on one particle species.
The resulting logarithmic negativity quantifies the entanglement between $\uparrow$ and $\downarrow$ particles on the two considered lattice sites.
What can be observed is a fast initial growth of entanglement during the flock formation where this growth is more prominent for shorter distances $d$.
The key observation, however, is that at long times $E_N^d(t)$ settles to a nonzero value independent of distance $d$((see Fig. \ref{fig:Log_neg_corr}C).
One explanation for this small nonzero value is the monogamy of entanglement.
If a subsystem $A$ is entangled strongly with another one $B$, it cannot be strongly entangled with any third party.
Conversely, if $A$ is entangled equally strongly with many others (here given by the different distances $d$) then the entanglement has to be shared accordingly with a corresponding smaller value.
The most important consequence is that the flocking steady state carries quantum entanglement.
Although the numerical methods used here to solve the Lindblad equation via stochastic unraveling are very powerful, we can access only reduced density matrices of small subsystems.
For the future it would therefore be of most interest to study entanglement of larger subsystems which promise even more striking quantum signatures.
}
\section{Classical analogue of the quantum flocking model}
In order to further argue, that the quantum model studied in the main text exhibits a flocking phase, we study here for completeness a classical analogue.
Concretely, we consider again a one-dimensional chain occupied by two species of particles $\sigma=\uparrow, \downarrow$.
As for the quantum model, we assume an excluded volume for these particles in that a given lattice site cannot be occupied by two particles of the same type.
The dissipative processes of the Lindblad master equation exhibit a straightforward classical analog.
Concerning the hopping of particles, with a direction depending on their internal degrees of freedom $\sigma$, we consider jump probabilities $P_{\mathcal{M}_{l\uparrow}}=(n_{l\uparrow} + 1)(n_{l+1\uparrow}-1)$ and $P_{\mathcal{M}_{l\downarrow}}=(n_{l+1\downarrow}+1)(n_{l\downarrow}-1)$.
Here, $n_{l\sigma}$ denotes the occupation on lattice site $l$ with a particle of species $\sigma$ and $m_{l} = n_{l\uparrow}-n_{l\downarrow}$ the magnetization.
As in the quantum case, this is the process which makes the particles active.
Further, for the alignment process we define a jump probability $P_{{\cal A}_{l\sigma}}= \mathrm{exp}(-K/(2r){m}_l\sum_{|j|=1}^{r}{m}_{l+j})(n_{l\sigma}+1)(n_{l\bar{\sigma}}-1)$ with $K$ denoting the alignment parameter and $r$ defining the interaction radius, as in the quantum model. \changed{Additionally, similar to how coherent spin flips in the quantum model, driven by the Hamiltonian, serve as a noise-like mechanism, this classical model introduces an extra jump process, described by $P_{\gamma}=\gamma(n_{l\sigma}+1)(n_{l\bar{\sigma}}-1)$, which plays a comparable role. This addition enhances the dynamic complexity and further bridges the behavior of the classical model with that of its quantum counterpart.}
We solve the resulting classical dynamics numerically by means of a synchronous update algorithm as in Ref.~[S3]

\begin{figure}[t]
\centering
\includegraphics[width=0.6\columnwidth]{Fig6.pdf}
\caption{
\textbf{Classical analogue of the quantum flocking problem.} Dynamics of magnetization fluctuations \(M^2(t)\) for the alignment parameter \(K=0.5\) in \textbf{A} and \(K=3.5\) in \textbf{B}. \changed{Real-space magnetization correlations for \(K=0.5\) in \textbf{C} and \(K=3.5\) in \textbf{D}. In all plots, the noise amplitude is \(\gamma=0\).}
}
\label{fig:order_t-classic}
\end{figure}

For the initial condition we choose a state with a vanishing magnetization in a specific  configuration, where particles are placed on the chain in $\uparrow$-and-$\downarrow$ pairs, occupying a quarter of the chain's $L$ sites and leaving the remaining part of the chain unoccupied. For a spin chain with L=8 sites, the initial state is given by $\{ \uparrow\downarrow, \uparrow\downarrow, 0, 0, 0, 0, 0, 0 \}$.
In order to detect the long-range ordered flocking phase in the present classical model we study the order parameter fluctuations:
\begin{equation}
 \langle M^2 (t)\rangle=\frac{1}{L^2}\sum_{lj} \langle m_l(t) m_j(t) \rangle~,  
\end{equation}
where $\langle...\rangle$ denotes the average over different trajectories,
$m_l=n_{l\uparrow}-n_{l\downarrow}$, and $n_{l\uparrow(\downarrow)}$ represent the number of up(down) spins at site $l$.
For a system with long-range order, we have that $M^2>0$ whereas for a disordered phase $M^2 \to 0$ in the thermodynamic limit.

\begin{figure}[t]
\centering
\includegraphics[width=0.4\columnwidth]{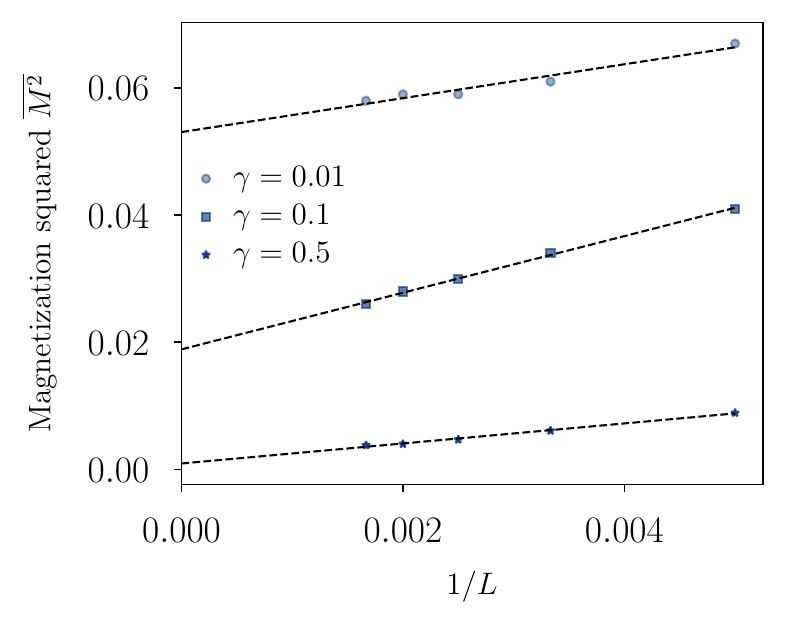}
\caption{
\textbf{Classical analogue of the quantum flocking problem.}  Long time average of Magnetization squared as a function of $1/L$ for different values of noise $\gamma$. We have $K=3.8$.
}
\label{fig:mag-classic_noise}
\end{figure}

In Fig.~\ref{fig:order_t-classic}(A) and (B), we show the numerically obtained  evolution of $M^2$ for two different values of the alignment parameter, $K=0.5$ and $K=3.5$, as well as for different system sizes $L$.
We observe that in both cases the order parameter eventually reaches a non-zero value in the long-time limit.
Importantly, however, for $K=0.5$ the value of $M^2$ becomes smaller as the system size $L$ increases, whereas for $K=3.5$ we find that $M^2$ remains constant.
This suggests that this classical model exhibits a flocking phase at sufficiently large values of the alignment parameter $K$. \changed{ Additionally, we demonstrate how the long-time average of \(M^2\) depends on the system size under non-zero noise conditions, indicating long-range order even in the presence of noise; see Fig.~\ref{fig:mag-classic_noise}.}

\changed{We extend our analysis to examine spatial correlation functions, which serve as an additional metric for assessing long-range order within the system. Specifically, we focus on the real-space magnetization correlations defined as
\begin{equation}
    m_{d} = L^{-1}\sum_l \langle \hat m_l \hat m_{l+d} \rangle~.
\end{equation}
These correlations are plotted as a function of the separation distance $d$ in Fig.~\ref{fig:order_t-classic}(C) and (D), for $\gamma=0$ and $K=0.5, 3.5$, respectively.
}

\section{Alternative dissipative alignment processes}

\begin{figure}[t]
\centering
\includegraphics[width=0.6\columnwidth]{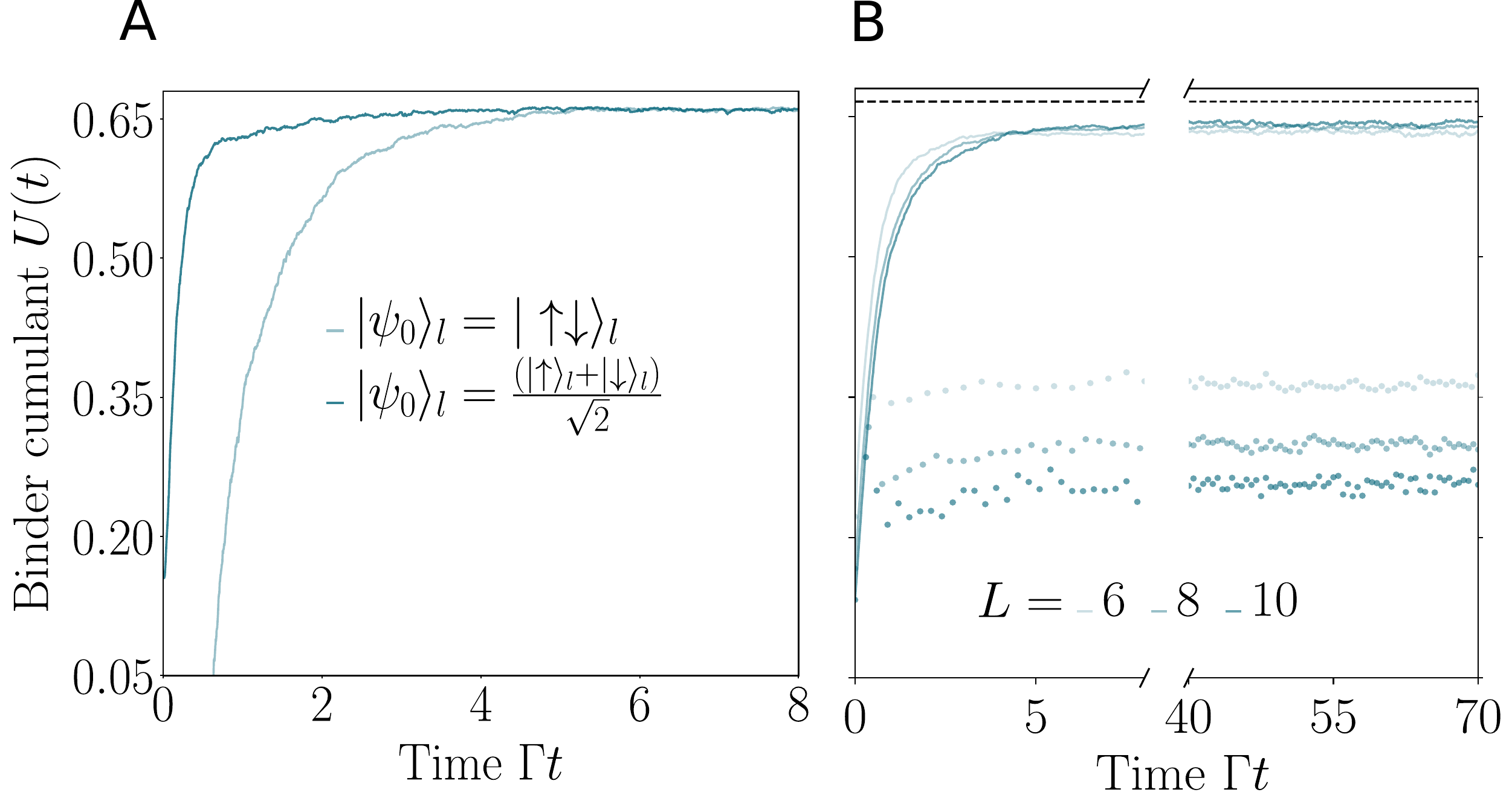}
\caption{
\textbf{Quantum dynamics for alternative dynamical setups.} \textbf{A} Dynamics of the Binder cumulant $U(t)$ for two types of initial product states, as indicated by the legend, converging to the same asymptotic steady state value. The numerical data has been obtained for a system size $L=8$, quantum amplitude $g/\Gamma=1/2$, and alignment parameter $K=3.8$. \textbf{B} Binder cumulant $U(t)$ obtained for two different unravelings of the Lindblad master equation and $g/\Gamma=0.2,2.0$ and $K=2.0$.
}
\label{fig:compare_initial}
\end{figure}

As discussed in the main text we have also explored alternative versions of dissipative alignment processes with quantum jump operators:
\begin{equation}
    \quad\hat{\mathcal{A}}_{l\sigma}=\hat{c}^{\dagger}_{l\sigma}\hat{c}_{l\overline{\sigma}} \hat P_l \, ,
\end{equation}
by utilizing different variants of $\hat P_l$ leading to different kinds of conditional particle-flip processes.
This includes:
\begin{enumerate}
    \item $\hat P_l= \mathrm{exp}(-K/(2r)\hat{m}_l\sum_{|j|=1}^{r}\hat{m}_{l+j})$, which is utilized in the main text.
    \item $\hat P_l= 1-K/(2r)\hat{m}_l\sum_{|j|=1}^{r}\hat{m}_{l+j}$, which can be viewed as the linearized version of the previous one.
    \item $\hat P_l=\sum_\sigma \hat P_{l\sigma}2^{-1}$ with $\hat P_{l\sigma}=\hat m_l (1+\sigma \hat m_l) \delta(\sum_{|j|=1}^{r}\hat{m}_{l+j} +\sigma M_0)$. This version induces jumps just when the surrounding magnetization $\sum_{|j|=1}^{r}\hat{m}_{l+j}$ matches precisely some predefined value $M_0$.
\end{enumerate}
In Fig.~\ref{fig:compare_initial}, we show the numerical data obtained for all these three cases.
As one can see, in all variants we find evidence for a quantum flocking phase.
These numerical observations are consistent with the coarse-grained analysis of the Lindblad master equation in the main text, where in the end the sole role of the $\hat P_l$ operator is to provide some nonlinear terms in the magnetization in the respective Taylor expansion. 

\section{Initial conditions}
In this section, we aim to provide evidence that the emergence of quantum flocks in our model is not specific to the precise details of the initial condition.
For that purpose, we show in Fig.~\ref{fig:compare_initial} the dynamics corresponding to two initial states: $|\psi_0\rangle = \otimes_{l=1}^{N} |\psi_0\rangle_l \otimes_{l=N+1}^L \ket{0}_l$ with $|\psi_0\rangle_l = 2^{-1/2}(\ket{\uparrow}_l + \ket{\downarrow}_l)$ and $|\psi_0\rangle_l = \ket{\uparrow\downarrow}_l$. 

In Fig.~\ref{fig:compare_initial} we show the evolution of the Binder cumulant $U(t)$ for these two initial conditions for a system of size $L=8$, alignment parameter $K=3.8$ and  quantum amplitude $g/\Gamma=1/2$.
While the dynamics of the Binder cumulant differs on short times, as expected for different initial conditions, both time evolutions reach the same asymptotic long-time steady state value of $U(t)$.
\section{Experimental Realization}
In the following we discuss how the elementary building blocks of the quantum dynamics described in the main text may be realized 
in an experimental setup.
We consider the directed motion induced by the quantum jump operators $\mathcal{M}_{l\sigma}$
separately from the alignment dynamics $\mathcal{A}_{l\sigma}$.
~\\[.2cm]
\noindent
\textbf{Directed Motion.}

In order to realize the directed motion, we take as the starting point a $U(1)$ lattice gauge theory in the form of a so-called quantum link model~[S4,S5].
Concretely, we consider matter coupled to a spin-$1/2$ gauge degree of freedom on the link with a minimal coupling.
For a single bond this reduces to the following Hamiltonian~[S4,S5]:
\begin{equation}
    H = \omega (c_1^\dag S^+ c_2 + \mathrm{h.c.} )
\end{equation}
with $\omega$ denoting the amplitude of that process.
Here, $S^{+}$ and $S^-$ are the usual spin-$1/2$ ladder operators and $c_{1/2}$ denote some annihilation operators.
In analogy to the model in the main text we consider hard-core bosons.
In the following, we will introduce an abbreviation $B^\dag = c_1^\dag c_2$ so that we work with the general Hamiltonian:
\begin{equation}
    H=\omega(S^+ B^\dagger + S^- B) \ .
\end{equation}
In order to generate the dissipative directed motion we subject this Hamiltonian to additional fast dissipative dynamics in the form of a decay of the gauge degree of freedom.
This corresponds to a Lindblad master equation of the form:
\begin{align}
 \frac{d\rho}{d t} = -i [H,\rho] + \gamma S^- \rho S^+
 - \frac{\gamma}{2} \{ S^+ S^-, \rho\} \, ,
\end{align}
with $\gamma$ the rate of the decay process, which we will choose as $\gamma \gg \omega$.
The underlying mechanism is that the fast spin decay ensures that - although  the interaction Hamiltonian is Hermitian - the process $S^+ B^\dagger$ is favored by the dissipative system as compared to $B S^-$.
We proceed to adiabatically eliminate the fast dynamics related to the spin decay.
To this end, we follow the construction shown in Ref.~[S9] but we note that in Ref.~[S9] an internal degree of freedom is eliminated and here we eliminate an external degree of freedom.
We thus consider the projector $P = \ket{\up}\bra{\up} \otimes \mathds{1}$
and find the altered Hamiltonian
\begin{align}
 \bar{H} = (\mathds{1} - P) H (\mathds{1} - P) = 0.
\end{align}
The quantum jump operator $Q = S^- \otimes \mathds{1}$ is altered as
\begin{align}
 \bar{\Gamma}\bar{Q} = (\mathds{1} -P) Q H (\mathds{1} - P) =
 \omega \ket{\down}\bra{\down} B^\dagger
\end{align}
with $\bar{\Gamma} = \epsilon \gamma$. The slow dynamics described by the 
slow component $\varrho_s = (\mathds{1} - P) \rho (\mathds{1} - P) +  
 Q \rho Q^\dagger$ of the full density matrix $\rho$
is hence revealed to follow the Lindblad master equation
\begin{align}
 \frac{d \varrho_s}{d t} = & 2\frac{\omega^2}{\gamma}
 \left(
 2\ket{\down}\bra{\down} B^\dagger \varrho_s\ket{\down}\bra{\down} B - \right. \nonumber \\
 & \left. - \ket{\down}\bra{\down} BB^\dagger \varrho_s - \varrho_s \ket{\down}\bra{\down} BB^\dagger
 \right)
\end{align}
Finally, we take the partial trace over the spin $1/2$ degree of freedom, i.e.,
$\varrho = \operatorname{tr}_{1/2} \varrho_s$ and we obtain the Lindblad master equation
\begin{align}
 \frac{d \varrho}{d t} = \frac{2\omega^2}{\gamma}\left( 2B^\dagger \varrho B
 -B B^\dagger - \varrho B  B^\dagger \right)
\end{align}
Recalling that $B^\dag = c_1^\dag c_2$ we find that this effective dynamics is exactly of the form of the directed motion used in our model in the main text.
Here, we have discussed an experimental realization for a single bond.
The extension to a full chain is straightforward by starting with a lattice gauge theory on that chain and imposing strong independent decay channels for all the gauge spins.
For the final realization of the directed motion in the considered quantum model, it would furthermore be necessary to implement the described dynamical process for two types of particles leading to the two-species scenario studied in the main text.
\begin{figure}
    \centering
    \includegraphics[width=.6\columnwidth]{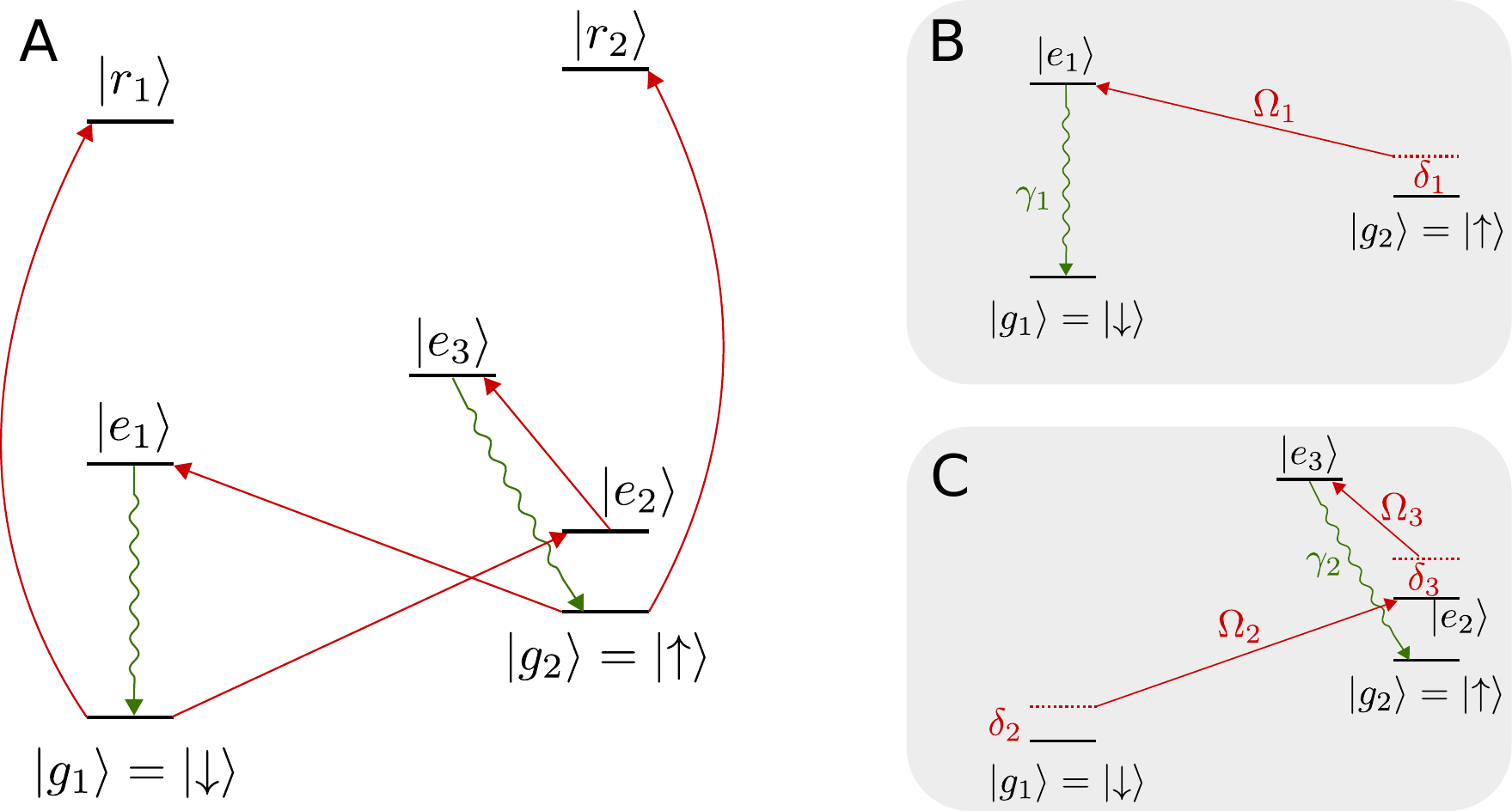}
    \caption{{\bf Level scheme.} 
    Panel A shows the full level scheme of a Rydberg atom that can encode both particle species $\ket{\up/\down}$ in the states $\ket{g_{1/2}}$. These are coherently coupled to highly excited Rydberg states $\ket{r_{1/2}}$ to implement an alignment dynamics that is akin to the one used in the main text, exploiting the Rydberg blockade mechanism. Panel B and C show the dissipative spin flip processes through coherent coupling to excited states (red) that decay dissipatively (green).}
    \label{fig:levels}
\end{figure}

~\\[.2cm]
\noindent
\textbf{Alignment dynamics.}
In a next step we discuss how a the alignment process for the quantum flock might be realized in a system of Rydberg atoms.
For that purpose we again consider just a single building block consisting of one atom with two internal states $\uparrow$ and $\downarrow$, for which we implement an effective dissipative spin-flip conditioned on an environment.

Fig.~\ref{fig:levels} illustrates the level scheme we consider to realize the alignment dynamics.
The up to down flip ($\ket{\up} \to \ket{\down}$) in particular is highlighted in 
Fig.~\ref{fig:levels}B. The three-level system involved in this transition is described by the density matrix $\rho$ that follows the Lindblad master equation
\begin{align}
    \frac{d\rho}{dt} = -i [H_1,\rho] +
    \gamma_1\left(2L\rho L^\dagger - L^\dagger L \rho - \rho L^\dagger L\right)
\end{align}
with the quantum jump operator $L = \ket{g_1}\bra{e_1}$ modelling the dissipative 
$\ket{e_1} \to \ket{g_1}$ transition and the Hamiltonian
$H_1 = \delta_1 \ket{g_2}\bra{g_2}+ \Omega_1 \ket{g_2}\bra{e_1}+ \Omega_1^* \ket{e_1}\bra{g_2}$
describing the coherent driving of the $\ket{g_2} \to \ket{e_1}$ transition.
For $\gamma_1 \gg |\delta_1|,|\Omega_1|$ we may adiabatically eliminate the state $\ket{e_1}$ following, e.g., Ref.~[S9]. The effective Hamiltonian
of the reduced system is thus given by
\begin{align}
    H_1^{\rm (eff)} = (\mathds{1} - \ket{e_1}\bra{e_1}) H_1 (\mathds{1} - \ket{e_1}\bra{e_1}) = \delta_1 \ket{g_2}\bra{g_2}
\end{align}
and the effective quantum jump operator $L^{\rm (eff)}$ reads
\begin{align}
    L^{\rm (eff)} = \frac{1}{\gamma_1}(\mathds{1} - \ket{e_1}\bra{e_1}) L H (\mathds{1} - \ket{e_1}\bra{e_1})
    = \frac{\Omega^*}{\gamma_1} \ket{g_1}\bra{g_2}
\end{align}
Writing $\sigma^- = \ket{g_1}\bra{g_2}$ and $\sigma^+ = (\sigma^-)^\dagger$ we 
find the effective Lindblad master equation by simply replacing 
$H_1 \to H_1^{\rm(eff)}$ and $L \to L^{\rm (eff)}$ in the original 
Lindblad equation, viz.,
\begin{align}
    \frac{d\varrho_s}{dt} = &  -i [H_1^{\rm (eff)},\rho_s] + \nonumber \\
     & + 2 \frac{|\Omega_1|^2}{\gamma_1} \left(2\sigma^-\rho \sigma^+
    - \sigma^+ \sigma^- \rho - \rho \sigma^+ \sigma^-\right)
\end{align}

Similarly, we may proceed for the down to up flip ($\ket{\down} \to \ket{\up}$)
illustrated in Fig.~\ref{fig:levels}C. Here, we need to eliminate the excited
states $\ket{e_2}$ and $\ket{e_3}$.
We observe that the state $\ket{g_2}$ does not couple coherently to the remainder of
the system. Hence, the remainder forms a subsystem reminiscent to a standard 
Lambda system, only that the coherently coupled state is in the middle of the spectrum.
This detail merely reverses the sign of a single detuning such that the Hamiltonian
may be written as
\begin{align}
H_{2} = \delta_2 \ket{g_1}\bra{g_1} - \delta_3 \ket{e_3}\bra{e_3} + \Omega_2 \ket{g_1}\bra{e_2} + \nonumber \\ + \Omega_2 \ket{e_2}\bra{g_1}
+ \Omega_3 \ket{e_2}\bra{e_3} + \Omega_3^* \ket{e_3}\bra{e_2}
\end{align}
The Lindblad equation dissipatively coupling the state $\ket{g_2}$ then reads
\begin{align}
    \frac{d \rho}{d t}= - i [H,\rho] + \frac{\gamma_2}{2} \left( 
    2 L \rho L^\dagger - L^\dagger L \rho - \rho L^\dagger L \right)
\end{align}
with the quantum jump operator $L = \ket{g_2}\bra{e_3}$. Now, we first eliminate the
state $\ket{e_3}$ for $\gamma_2 \gg |\Omega_3|,|\delta_3|$ similar to the 
opposite spin flip discussed above. This yields a new damping constant 
$\gamma' = 2 |\Omega_3|^2/\gamma_2$ and for $\gamma' \gg |\Omega_2|, |\delta_2|$ we 
may adiabatically eliminate the state $\ket{e_2}$. The elimination condition is 
generally met since $|\Omega_2| \ll \gamma' \sim |\Omega_3|^2/\gamma_2 \ll \gamma_2$.
This procedure yields the effective Lindbladian
\begin{align}
    \frac{d\tilde{\varrho}_s}{dt} = &  -i [H_2^{\rm (eff)},\tilde{\rho}_s] + \nonumber \\
    +& \frac{|\Omega_2|^2}{|\Omega_3|^2}\gamma_2 \left(2\sigma^- \rho \sigma^+
    - \sigma^+ \sigma^- \rho - \rho \sigma^+ \sigma^- \right)
\end{align}
with the effective Hamiltonian
\begin{align}
   H_2^{\rm (eff)} = \delta_2 \ket{g_1}\bra{g_1} .
\end{align}
~\\[.2cm]
\noindent
{\bf Rydberg dressing.} 
Having established the dynamics to induce dissipative spin flips in ultracold atoms we now need to 
introduce a mechanism that is capable to (de)activate the spin flips depending on the surrounding 
magnetization, which would effectively realize the alignment process 3 discussed in the Supplement Sec.~II. To achieve this we dress both spin states respectively with a Rydberg state
$\ket{r_{1/2}}$, see Fig.~\ref{fig:levels}.
The key impact of this Rydberg dressing is to shift the energy levels $\delta_1$ and $\delta_2$ corresponding to the states $|\uparrow\rangle$ and $|\downarrow\rangle$ depending on the presence of other atoms.
Consider for instance a situation where only for a specific environment $\delta_1=\delta_2=0$, leading to the desired dissipative spin flip.
Additionally, let's assume that for any other environment $\delta_{1/2} \gg \gamma_{1/2}$, thus we can effectively neglect the dissipative process.
Summarizing:
\begin{itemize}
    \item the Rydberg dressing of $\ket{\down}$ with $\ket{r_1}$
    switches the $\ket{\down}\to\ket{\up}$ transition on/off.
    \item the Rydberg dressing of $\ket{\up}$ with $\ket{r_2}$
    switches the $\ket{\up}\to\ket{\down}$ transition on/off.
\end{itemize}
\begin{figure}[h]
\centering
\includegraphics[width=0.65\columnwidth]{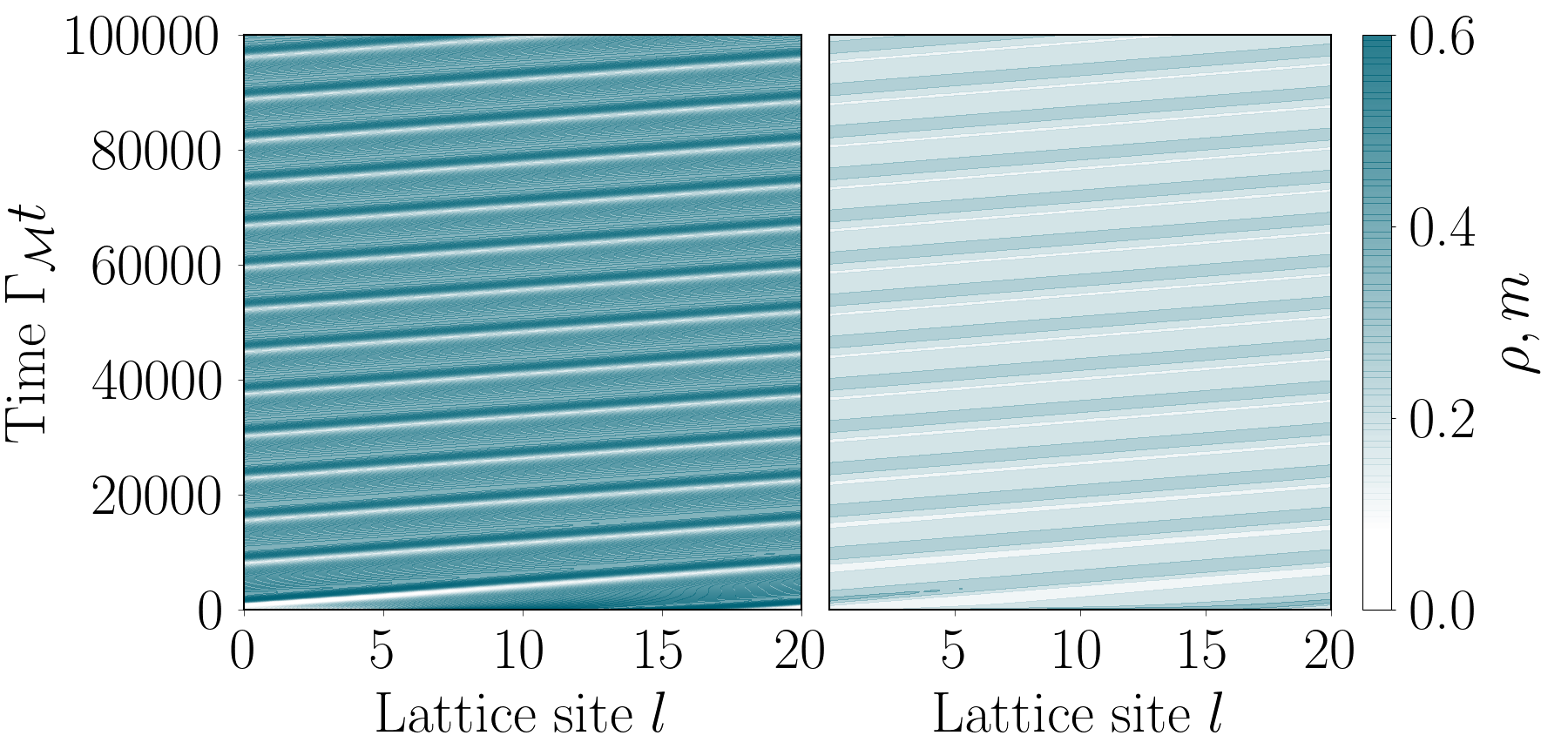}
\caption{
{\bf Dynamics of traveling wave patterns.} Dynamics of the density (left) and magnetization (right) for an initial cluster as a function of lattice site $l$ and time $t$ of the coarse-grained solution of the Lindblad-master equation obtained for parameters $L=20$, $K=4$, $\gamma_\rho=0.2$, $\gamma_m=0.6$, $\Gamma_\mathcal{A}/\Gamma_\mathcal{M}=0.1$ 
}
\label{fig:hydro_gauss}
\end{figure}
\section{Inhomogeneous solutions of coarse-grained dynamics}
Here, we provide details on inhomogeneous solutions of coarse-grained dynamics. We have seen, that the dynamical system allows for a collectively ordered phase if one includes fluctuations.
However, a key next step is to explore the stability of inhomogeneous solutions, which was also investigated in the context of classical flocks composed of self-propelled particles~[S6--S8].
Without loss of generality we consider the case $g=0$ and focus on Eqs.~(8) and~(9) in 
the limit $L\to\infty$. 
Similar to the homogeneous case, we ignore density-magnetization correlations such that Eq.~(8) is readily
rewritten in its continuum form with $x=al$ and $a$ the lattice spacing:
\begin{align}
\label{eq:rho_cont}
 \partial_t \rho = -\Gamma_\mathcal{M} \left( 2\partial_x ( \rho_x m_x) - \partial_x m_x 
 -\frac{1}{2}\partial_x^2 \rho_x \right)
\end{align}
with $\rho_x = \langle \hat \rho_l \rangle$ and $m_x = \langle \hat m_l \rangle$. 
Here, we explicitly neglect lattice 
constants ($a=1$) that differentiate finite differences from derivatives since we shall translate the 
continuum equations in a factorized form back to the lattice. The continuum form therefore solely serves
as a tool to identify relevant fluctuations.

\begin{figure}[t]
\centering
\includegraphics[width=0.65\columnwidth]{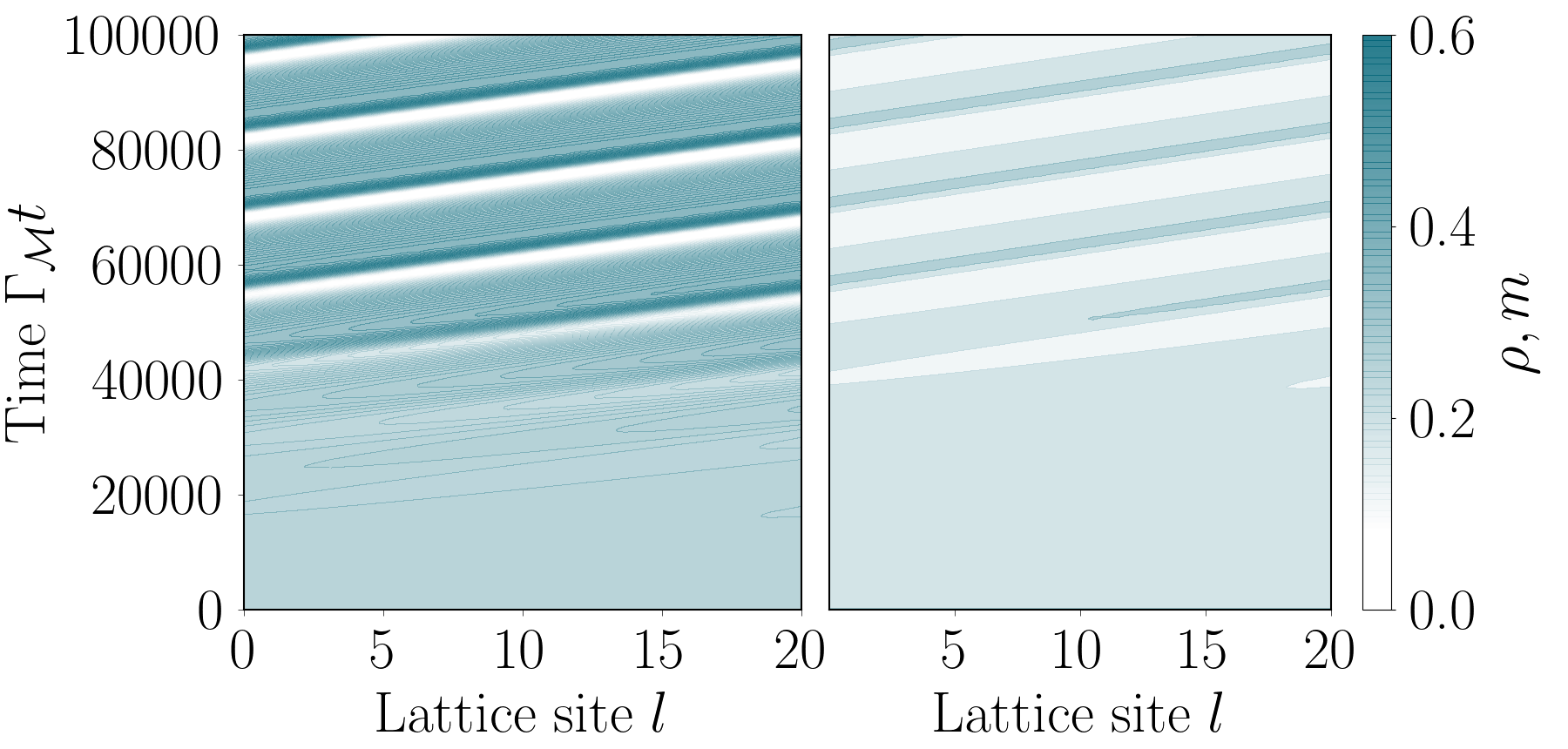}
\caption{
{\bf Emergence of traveling wave patterns.} Dynamics of the density (left) and magnetization 
profile (right) for an initial homogeneous state with weak noise. The homogeneous solution is unstable and inhomogeneous patters are formed after an initial 
transient regime.
Simulations parameters:
$L=20$,
$K=4$,
$\gamma_\rho = 0.2$,
$\gamma_m = 0.6$,
$\Gamma_M = 1$,
$\Gamma_F = 0.1$.}
\label{fig:hydro_noise}
\end{figure}

For the present inhomogeneous case we have to slightly adapt 
Eq.~(13) and introduce a mean magnetization $M=\langle \hat M^l \rangle$ which yields
\begin{align}
 \langle m_l P^\dagger_l P_l\rangle
 \simeq
 \langle m_l\rangle &- 2K \langle m_l^2\rangle M
+2K^2 \langle m_l^3\rangle M^2 .
\end{align}
The continuum form of Eq.~(9) then reads
\begin{align}
\label{eq:m_cont}
\begin{split}
  \partial_t m_x = &- \Gamma_\mathcal{M} \big( \partial_x \langle\rho^2_x\rangle-
 \partial_x \rho_x + \partial_x \langle m^2_x \rangle - \frac{\partial_x^2 m_x}{2}\big)\\
 &-2\Gamma_\mathcal{A} \big(
 m_x - 2K \langle m^2_x \rangle M + 2K^2 \langle m^3_x \rangle M^2
 \big).
 \end{split}
\end{align}
Similar to the analysis for the homogeneous solution, we consider the density and magnetization distribution 
as Gaussian fluctuating quantities, i.e., 
\begin{align}
 \langle X^2\rangle &= X^2 + \sigma_X^2,\\
 \langle X^3\rangle &= X^3 + 3\sigma_X^2 X,
\end{align}
for $X = \rho_x,m_x$. Furthermore, we express the variances $\sigma_X^2$ in terms of the density, i.e.,
$\sigma_X^2 \simeq \gamma_X \rho$, which closely follows the analysis performed in classical active Ising models~[S1,S2]. This renders Eq.~(\ref{eq:m_cont}) to be of the form
\begin{align}
\begin{split}
 \partial_t m_x = &
 -\Gamma_\mathcal{M} \big(  \partial_x \rho^2_x  
 - \partial_x\rho_x
 +  \partial_x m^2_x   - \frac{\partial_x^2 m_x }{2}\\
 &\qquad + [\gamma_\rho + \gamma_m] \partial_x\rho_x \big)\\[.4cm]
 &-2\Gamma_\mathcal{A} \big( m_x - 2K m^2_x M + 2K^2  m^3_x M^2\\
 &\qquad   -2K \gamma_m \rho_x M + 6 K^2 \gamma_m \rho_x m_x M^2
 \big).
 \end{split}
 \label{eq:m_cont_final}
\end{align}
Here, we have accounted for all relevant Gaussian fluctuations in the thermodynamic 
limit. 
We solve Eqs.~(\ref{eq:rho_cont}) and~(\ref{eq:m_cont_final}) numerically for different initial
states in Figs.~\ref{fig:hydro_gauss} and~\ref{fig:hydro_noise}.
Our numerical analysis of the coarse-grained description exhibits two main features.
When $\Gamma_\mathcal{M}=\Gamma_\mathcal{A}$, as we have for the exact numerics in the main text, we find that the homogeneous solution is the attractor of the dynamics.
Most importantly, however, we find that for $\Gamma_\mathcal{M}/\Gamma_\mathcal{A}=0.1 $ also stable inhomogeneous solutions exist in the form of traveling patterns.
In Fig.~\ref{fig:hydro_gauss} we consider an initial state
where the system only contains $\up$ particles and their 
density is distributed according to a Gaussian distribution $1/2 \exp(-0.5((x - L/2)/\sigma)^2)$ with $\sigma = 0.5$. 
We observe that the initial cluster is stable in time with a persistent wave
pattern visible in Fig.~\ref{fig:hydro_gauss}.

On top of this, we further find that in the considered regime the homogeneous solution actually becomes unstable, as we evidence in Fig.~\ref{fig:hydro_noise}.
There, we consider an initial condition with a homogeneous magnetization and density, i.e., $\rho = m = 0.25$, corresponding to a homogeneous flocking state of our quantum model, with a slight noise added on top of the homogeneous background. 
Concretely, we added to each site a uniformly distributed random number in $[-5\times 10^{-4},5\times 10^{-4}]$. We see that the system stays close to the 
homogeneous solution for some time, but eventually a traveling wave pattern emerges.


\begingroup
\small
\begin{enumerate}[label={[S\arabic*]},leftmargin=*,itemsep=1pt,parsep=0pt,topsep=4pt]

\item
A.~P.~Solon and J.~Tailleur, \href{https://doi.org/10.1103/PhysRevLett.111.078101}
{Phys.\ Rev.\ Lett.\ \textbf{111}, 078101 (2013).}

\item
A.~P.~Solon and J.~Tailleur, \href{https://doi.org/10.1103/PhysRevE.92.042119}
{Phys.\ Rev.\ E \textbf{92}, 042119 (2015).}

\item
R.~Singh and S.~Mishra, \href{https://arxiv.org/abs/1704.04041}
{arXiv:1704.04041.}

\item
S.~Chandrasekharan and U.~J.~Wiese, \href{https://doi.org/10.1016/S0550-3213(97)80041-7}
{Nucl.\ Phys.\ B \textbf{492}, 455 (1997).}

\item
M.~C.~Ba\~nuls \textit{et al.}, \href{https://doi.org/10.1140/epjd/e2020-100571-8}
{Eur.\ Phys.\ J.\ D \textbf{74}, 165 (2020).}

\item
E.~Bertin, M.~Droz, and G.~Gr\'egoire, \href{https://doi.org/10.1088/1751-8113/42/44/445001}
{J.\ Phys.\ A: Math.\ Theor.\ \textbf{42}, 445001 (2009).}

\item
C.~A.~Weber, F.~Th\"uroff, and E.~Frey, \href{https://doi.org/10.1088/1367-2630/15/4/045014}
{New J.\ Phys.\ \textbf{15}, 045014 (2013).}

\item
F.~Th\"uroff, C.~A.~Weber, and E.~Frey, \href{https://doi.org/10.1103/PhysRevX.4.041030}
{Phys.\ Rev.\ X \textbf{4}, 041030 (2014).}

\item
M.~Mirrahimi and P.~Rouchon, \href{https://doi.org/10.1109/TAC.2009.2015542}
{IEEE Trans. Autom. Control. \textbf{54}, 1325 (2009).}

\end{enumerate}
\endgroup